\pdfoutput=1
\documentclass[fleqn, usenatbib, usedcolumn, useAMS]{mnras}

\usepackage[T1]{fontenc}
\usepackage[british]{babel}
\usepackage{amsmath} 
\usepackage{amssymb} 
\usepackage{txfonts}
\usepackage{microtype}
\usepackage{booktabs}
\usepackage[pdftex]{graphicx} 

\hypersetup{pdfauthor={Philip D. Hall},
  pdftitle={Hydrogen in hot subdwarfs formed by double helium white dwarf mergers},
  pdfkeywords={binaries: close, stars: evolution, subdwarfs, white dwarfs},
  bookmarksnumbered=true}

\newcommand{\Msol}[0]{\rmn{M}_{\sun}}

\title[Hydrogen in HeWD+HeWD mergers]{%
 Hydrogen in hot subdwarfs formed by double helium white dwarf mergers}
\author[P.\ D.\ Hall \& C.\ S.\ Jeffery]{
Philip D.\ Hall$^{1}$\thanks{E-mail: pdh@arm.ac.uk (PDH); csj@arm.ac.uk (CSJ)} and C.\ Simon Jeffery$^{1,\,2}\footnotemark[1]$
\\
$^{1}$Armagh Observatory, College Hill, Armagh BT61 9DG, UK\\
$^{2}$School of Physics, Trinity College Dublin, College Green, Dublin 2, Ireland\\
}
\date{Accepted 2016 August 29. Received 2016 August 29; in original form 2016 May 25}

\begin{document}
\label{firstpage}
\maketitle
\volume{463}
\pagerange{2756--2767}
\pubyear{2016}

\begin{abstract}
Isolated hot subdwarfs might be formed by the merging of two helium-core white dwarfs.
Before merging, helium-core white dwarfs have hydrogen-rich envelopes and some of this hydrogen may survive the merger.
We calculate the mass of hydrogen that is present at the start of such mergers and, with the assumption that hydrogen is mixed throughout the disrupted white dwarf in the merger process, estimate how much can survive.
We find a hydrogen mass of up to about $2 \times 10^{-3}\,\Msol$ in merger remnants.
We make model merger remnants that include the hydrogen mass appropriate to their total mass and compare their atmospheric parameters with a sample of apparently isolated hot subdwarfs, hydrogen-rich sdBs.
The majority of these stars can be explained as the remnants of double helium white dwarf mergers.
\end{abstract}


\section{Introduction}\label{sec:intro}
A hot subdwarf is a member of one of the subdwarf B (sdB), helium sdB (He-sdB), subdwarf O (sdO) or helium subdwarf O (He-sdO) classes.
These stars are recognized and classified by properties of their spectra, such as the profiles and relative strength of hydrogen and helium spectral lines, or by detailed spectral analysis and comparison with model atmospheres \citep{2009ARA&A..47..211H, 2016PASP..128h2001H, 2013A&A...551A..31D}.
Such atmospheric modelling shows that these stars have effective temperatures, $T_{\rmn{eff}}$, between about $20$ and $80\,\rmn{kK}$ and logarithmic surface gravities, $\log_{10}(g/\rmn{cm}\,\rmn{s}^{-2}) \equiv \log g$, between about $5$ and $6.5$.
The sdB and He-sdB stars have smaller $T_{\rmn{eff}}$ than the sdO and He-sdO stars.
The helium abundance is often parametrized by the ratio of helium to hydrogen photospheric number densities, $y=n_{\rmn{He}}/n_{\rmn{H}}$.
Generally, sdBs and sdOs have $y \la 0.1$, and He-sdBs and He-sdOs have $y \ga 10$ \citetext{\citealp{2005A&A...430..223L}; \citealp{2007A&A...462..269S}; \citealp*{2012MNRAS.427.2180N}}.

The bulk of hot subdwarfs are believed to be stars with burning helium cores and inert hydrogen envelopes of low mass, their immediate progenitors or their shell-helium-burning progeny.
In particular, the sdBs are thought to be largely extended (or extreme) horizontal branch (EHB) stars: stars with helium cores of about $0.47\,\Msol$ and hydrogen envelopes of low mass, $\la 20 \times 10^{-3}\,\Msol$ \citep{1986A&A...155...33H}.
In the effective temperature--gravity plane the sdBs are concentrated in the region expected for the long-lived core-He burning phase between the ignition of central He (zero-age EHB, ZAEHB) and core-He exhaustion (terminal-age EHB).
Core-He-burning stars with more or less massive He-rich cores ($\ga 0.33\,\Msol$) and varying H-rich envelope masses can also have the right surface properties to be hot subdwarfs \citep{2002MNRAS.336..449H}.
Not all hot subdwarfs are shell- or core-He-burning stars.
For example, some hot subdwarfs are pre-helium white dwarfs, shell-H-burning stars with inert degenerate helium cores and low-mass hydrogen envelopes \citep{2003A&A...411L.477H}, and others are pre-carbon--oxygen white dwarfs \citep{2003ARA&A..41..391V}.

How are stars with these structures formed?
A large fraction, about $2/3$, of hot subdwarfs have companions \citep{2001MNRAS.326.1391M} on short-period orbits (orbital period $P \la 10\,\rmn{d}$).
Systems of this type are explained with canonical models of the evolution of binary star systems: the hot subdwarf is a star which had nearly all of its envelope removed in a common-envelope (CE) phase and also ignited helium in its core.
Similarly, a few hot subdwarfs are found to have companions on long-period orbits \citep{2012MNRAS.421.2798D,2013A&A...559A..54V}.
These systems are also explained in the normal evolution of binary systems, but with the hot subdwarfs having had most of their envelopes removed in a phase of Roche lobe overflow \citetext{\citealp*{1976ApJ...204..488M}; \citealp{2002MNRAS.336..449H}; \citealp{2009A&A...503..151Y}}; although their eccentricity may be a puzzle.
We have more uncertain understanding of the formation of the isolated hot subdwarfs, those without apparent companions.
There have been several suggested channels through which such stars could form, so the problem is to find the extent to which each of these contributes -- if at all -- to the observed population. 
A hot subdwarf phase is not part of the evolution of a single star, unless one assumes that mass-loss is enhanced during its first red giant phase \citep{1993ApJ...407..649C, 1996ApJ...466..359D}.
On the other hand, standard evolutionary models of binary systems may produce isolated hot subdwarfs through various types of mergers of the two stars.
We discuss other proposed channels in Section~\ref{sec:discussion}, but our focus in this paper is on the channel involving the merging of two helium-core white dwarfs \citep{1984ApJ...277..355W, 1986ApJ...311..753I}.
The formation of a hot subdwarf through this channel begins with the formation of a short-period detached pair of helium-core white dwarfs.
Such systems were predicted theoretically \citep{1984ApJ...277..355W} before they were first observed \citep*{1988ApJ...334..947S} and several tens of candidates have since been discovered \citetext{\citealp*{1995MNRAS.275..828M}; \citealp{2005A&A...440.1087N}; \citealp{2011ApJ...727....3K, 2012ApJ...751..141K}; \citealp{2015ApJ...812..167G}}.
Some of the observed He+He systems have sufficiently short-period orbits ($P_{\rmn{orb}} \la 6\,\rmn{h}$) that, within a Hubble time, radiation of gravitational waves will decrease the separation to the point at which the lower mass white dwarf fills its Roche lobe and begins unstable mass transfer, ultimately resulting in the merging of the two white dwarfs to form a single He-rich star.
If this merger remnant is massive enough then helium is ignited, initiating a relatively long-lived ($\sim 100\,\rmn{Myr}$) phase as a core-He-burning object \citep{1990ApJ...353..215I}.
It is during this phase that the merger remnant would be a hot subdwarf.

Calculations of the remnants of double-white-dwarf (DWD) mergers show that this channel works from a theoretical perspective.
In fact, modelling of the outcome of DWD mergers, particularly of the mergers of two carbon--oxygen (CO) white dwarfs, has become a very active research area.
The activity is driven by the possibility that these CO+CO mergers are the progenitors of Type Ia supernovae \citep{1984ApJS...54..335I, 1984ApJ...277..355W}, so few studies have focused on the He+He DWD mergers of interest to this paper.
Generally, for mergers of white dwarfs of all core compositions, hydrodynamical simulations show that if detonation is avoided in the initial phases of a merger then one white dwarf is disrupted to form a disc-like structure around a cold core and hot envelope \citep{2014MNRAS.438...14D}.
However, the evolution after this point, the immediate post-disruption phase, is uncertain (see Section~\ref{sec:method}).
There have therefore been various methods and models used to calculate the evolution of merged white dwarfs \citep{1990ApJ...353..215I, 2000MNRAS.313..671S, 2012MNRAS.419..452Z}. 
While these models confirm that a He+He DWD merger can produce a He-dominated object that burns helium in its core for about $100\,\rmn{Myr}$, none of these studies has explicitly included the hydrogen which can dominate the envelopes (with masses $\la 10 \times 10^{-3}\,\Msol$) of the two white dwarfs in the start-of-merger configuration.
Thus these previous models have He+He DWD mergers producing He-rich hot subdwarfs: He-sdB and He-sdO stars.
The mass of hydrogen in the merger remnant is important in dictating its surface properties: its effective temperature, surface gravity and whether the surface is H- or He-rich.
Generally, the addition of a hydrogen envelope to a He main-sequence (MS) star decreases its effective temperature and surface gravity \citep{1961ApJ...133..764C, 1967ZA.....65..226G}. 
Previous work has ignored hydrogen because the high temperatures during the merger could be expected to lead to its destruction.
However, it is our intention in this paper to investigate the implications of the possibility that the hydrogen is distributed throughout the disrupted white dwarf and thus not heated to sufficiently high temperatures for burning \citep{1986ApJ...311..753I, 1990ApJ...353..215I}.
In the post-merger phase diffusion would then act to bring any surviving hydrogen to the surface of the remnant and form a subdwarf with a H-rich envelope.
In modelling the hot subdwarf population, \citet{2002MNRAS.336..449H, 2003MNRAS.341..669H} did include H-rich envelopes in their DWD merger remnants, although this was artificially added to H-free remnants.
They assumed a uniform distribution between $0$ and $10^{-3}\,\Msol$ for the H-rich envelope mass.
It was necessary to make this assumption for their model populations to show the spread in effective temperature observed.
However, their choice of an upper limit of $10^{-3}\,\Msol$ for the H-rich envelope mass was not justified.
It is not clear, for example, if this is reasonable for mergers of the highest mass He white dwarfs, which are expected to have lower mass envelopes \citep{1998A&A...339..123D}.
There is thus ample motivation to more carefully investigate hydrogen in He+He DWD mergers, and particularly to make use of recent simulations of the disruption phase of the merger \citep{2014MNRAS.438...14D}.

In this work we ask how much hydrogen can survive a He+He DWD merger to the point of core He burning.
By focusing on realistic configurations at the start of a merger, we estimate the mass of hydrogen and other nuclides that are present and survive the merger.
We use published simulations of the disruption of the less massive WD to estimate the mass of hydrogen remaining after this phase of the merger.
With some assumptions we use these estimates to predict the range of atmospheric properties -- the effective temperature, surface gravity and composition -- of the remnants of He+He DWD mergers in long-lived burning phases.
By doing so we consider the question of to what extent hot subdwarfs with H-rich surfaces can be explained as the progeny of DWD mergers.

In Section~\ref{sec:tools} we discuss the 1D stellar-evolution code we use,  \textsc{mesa/star}.
In Section~\ref{sec:method} we describe how we model merger remnants, starting with the discussion of a particular case and ending with a general method.
In Section~\ref{sec:sample} we choose a sample of isolated hot subdwarfs to compare to our models.
In Section~\ref{sec:results} we present the results of applying our method to a large set of start-of-merger mass combinations.
In Section~\ref{sec:discussion} we discuss the relevance of these results to the question of how isolated hot subdwarfs are formed.
In Section~\ref{sec:conclusion} we give our conclusions.

\section{\textsc{mesa/star} stellar evolution code}\label{sec:tools}
The models we produce here rely on the \textsc{mesa/star} stellar-evolution code from a recent release of the open-source Modules for Experiments in Stellar Astrophysics, \textsc{mesa} (revision 7624, released 2015-06-03; \citealp{2015ApJS..220...15P}).
In accordance with the \textsc{mesa} manifesto, we will make material necessary to reproduce our calculations available.
Model sequences produced by this code satisfy a standard set of equations describing one-dimensional quasi-static stellar evolution, full details of which can be found in the \textsc{mesa} instrument papers \citep{2011ApJS..192....3P, 2013ApJS..208....4P, 2015ApJS..220...15P} or by inspecting the freely available source code.
\textsc{mesa/star} allows for great flexibility in the choice of physics, so for those unfamiliar with the code, we summarize our choice of physics in the following paragraphs; for those familiar with the code, we summarize our deviations from default parameters in Appendix~\ref{sec:mesa_inlist}.

The equation of state and other thermodynamic quantities are computed using the default \texttt{mesa} option, which is based on a blend of the tables of \citet[OPAL]{2002ApJ...576.1064R}, \citet*[SCVH]{1995ApJS...99..713S}, \citet[HELM]{2000ApJS..126..501T} and \citet[PC]{2010CoPP...50...82P}.
The radiative opacity is that of the OPAL collaboration \citep{1996ApJ...464..943I}, including the effect of molecules, supplemented by the work of \citet{2005ApJ...623..585F} for low temperatures and of \citet{1976ApJ...210..440B} for pure electron-scattering at high temperatures.
We use the OPAL Type 2 opacity tables which allow for varying carbon and oxygen abundance.
The electron conduction opacity is taken from the work of \citet{2007ApJ...661.1094C}. 
The nuclear-reaction rates are those of \citet{1988ADNDT..40..283C} and the NACRE collaboration \citep{1999NuPhA.656....3A}. 
The enhancement of reaction rates by electron screening is included according to \citet{1973ApJ...181..457G} in the weak regime and \citet{1978ApJ...226.1034A} and \citet{1979ApJ...234.1079I} in the strong regime.
The rates of energy loss in neutrinos through the photo/pair, plasma, and bremsstrahlung processes are calculated using the formulae of \citet{1996ApJS..102..411I}.

In convective regions, regions that are unstable according to the Schwarzschild criterion, mixing is treated as diffusion with coefficients and temperature gradients computed according to the default mixing-length theory \citep{1968pss..book.....C}, with a mixing-length parameter $\alpha_{\rmn{MLT}}=2$.
In radiative regions, mixing occurs only by overshooting from convective boundaries, which is treated with the exponential overshooting formalism.
We choose an overshooting parameter $f_{\rmn{ov}} = 0.016$, achieved numerically with $f_0=0.004$ and $f=0.014$ \citep{2000A&A...360..952H}.

To treat nuclear reactions, our models use the \texttt{o18\_and\_ne22.net} network, which includes $10$ nuclides: $^{1}\rmn{H}$, $^{3}\rmn{He}$, $^{4}\rmn{He}$, $^{12}\rmn{C}$, $^{14}\rmn{N}$, $^{16}\rmn{O}$, $^{18}\rmn{O}$, $^{20}\rmn{Ne}$, $^{22}\rmn{Ne}$ and $^{24}\rmn{Mg}$.
When evolving from the pre-MS, models initially have solar-scaled initial abundances with the \citet[GS98]{1998SSRv...85..161G} mixture, except for metals which are not in the network; these have their abundances added to that of the heaviest nuclide, $^{24}\rmn{Mg}$.
We choose opacity tables with the same metals mixture.
Models in the pre-MS phase have zero-age helium mass fraction $Y_0=0.24 + 2Z_0$, where $Z_0$ is the zero-age metallicity.
We define the outer boundary of the helium core to be the point at which the hydrogen mass fraction $X(^1\rmn{H})=0.1$.
The surface boundary conditions are chosen to match an Eddington grey atmosphere at an optical depth $\tau = 2/3$.

\section{Method}\label{sec:method}
To explain our method of computing the mass of hydrogen that can survive a double helium white dwarf merger, we consider a specific case.
We discuss the structure of the two white dwarfs at the start of the merger and how we use this information and the results of dynamical merger simulations to find the surviving hydrogen mass.
The start-of-merger conditions for DWD mergers are usually parametrized in terms of the masses of the stars and the dominant contribution to their core composition -- helium (He), carbon and oxygen (CO), or oxygen, neon and magnesium (ONeMg).
As an example of the merger of two helium-core white dwarfs (a He+He DWD merger) we consider a $0.3$+$0.2\,\Msol$ case, which we choose because the mass of the merger remnant in this case is expected to be close to the canonical hot subdwarf mass.
We call the more massive white dwarf WD1 ($0.3\,\Msol$) and the less massive white dwarf WD2 ($0.2\,\Msol$).

\subsection{Pre-merger evolution}\label{sec:pre_merger_evolution}
Before merging, the system is most likely a short-period detached He+He DWD binary system in which the separation decreases by gravitational wave radiation.
Provided the separation at the beginning of this detached phase is sufficiently small ($P_{\rmn{orb}} \la 6\,\rmn{h}$), the stars are brought sufficiently close within a Hubble time for WD2 to fill its Roche lobe in a binary system with $P_{\rmn{orb}}\approx 4\,\rmn{min}$ \citep{1967AcA....17..287P, 1979AcA....29..665T}.
Our aim is to find the structures, particularly the H-rich envelope masses, of the white dwarfs at this point, the start of the merger.
To this end we now discuss the formation and evolution of a short-period detached DWD system and how the past evolution dictates the start-of-merger configuration.

A short-period detached $0.3$+$0.2\,\Msol$ He+He DWD system could be formed from a zero-age MS binary system through a number of qualitatively different channels, generally involving two mass-loss episodes, each of which removes the envelope of one star to expose a helium core and a low-mass hydrogen envelope, the progenitor of a WD \citetext{\citealp{1987fbs..conf..401I}; \citealp*{1996MNRAS.279...88S}; \citealp{1998MNRAS.296.1019H}; \citealp{2001A&A...365..491N}}.
Originally, the most likely channels were thought to be (a) a stable phase of Roche lobe overflow then a CE phase, the RLOF+CE channel, and (b) a CE phase then another CE phase, the CE+CE channel \citep{1984ApJ...277..355W,1984ApJS...54..335I}.
More detailed studies of observed He+He DWDs concluded that it is difficult to explain the first mass-loss episode as a CE phase \citetext{\citealp{2000A&A...360.1011N}; \citealp{2005MNRAS.356..753N}; \citealp*{2006A&A...460..209V}} and that it may instead be a phase of non-conservative Roche lobe overflow \citep{2012ApJ...744...12W}.
RLOF+RLOF or CE+RLOF channels may also be possible \citep{1996MNRAS.279...88S} but they have not been included in recent population models.
DWDs with nearly equal masses may have evolved through a single mass-loss episode, double-core CE phase in which both stars are evolved at the onset \citep{1995ApJ...440..270B, 2001A&A...365..491N}.
There is unlikely to be a unique set of zero-age conditions associated with a given pair of start-of-merger masses, so by giving only the masses and core compositions at the start of the merger we must represent the remnants of a range of zero-age conditions.

\begin{figure}
  \centering
  \includegraphics[width=84mm]{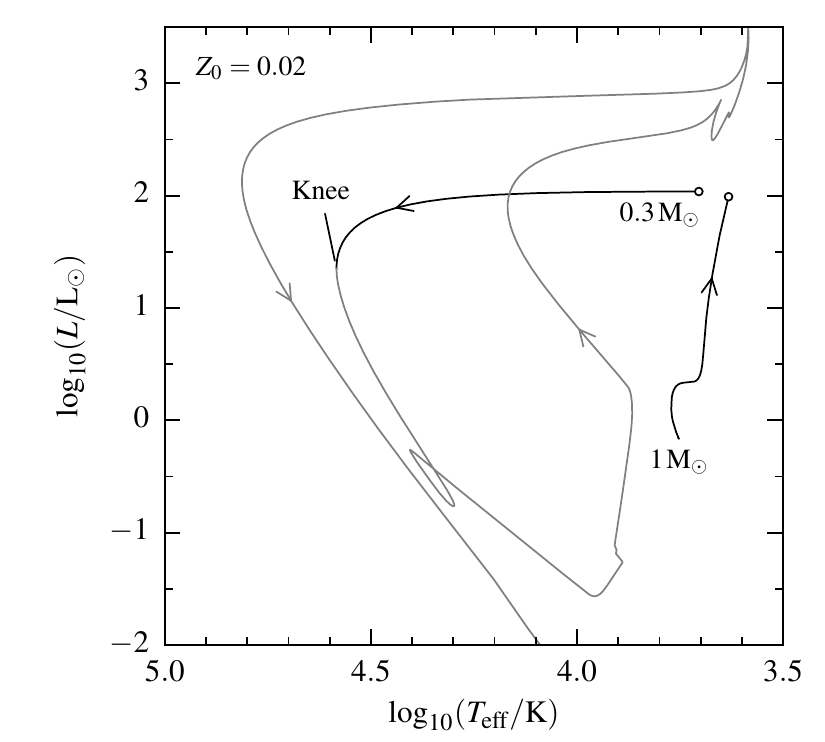}
  \caption{
Evolution in the effective temperature--luminosity plane of a $Z_0=0.02$, $M_0=1\,\Msol$ star from the zero-age MS until it becomes a red giant with a $0.28\,\Msol$ helium core, and then after it has become a $0.3\,\Msol$ mass-loss remnant.
Evolution before the knee stage is shown in black.
Evolution after the knee stage, when the star undergoes a shell flash, is shown in light grey.}
  \label{fig:Teff-L_HeWD}
\end{figure}

Assume first that our example $0.3$+$0.2\,\Msol$ system has evolved through the CE+CE evolution channel and that WD1 is the descendant of a low-mass star, star 1, that had mass $M_{10}=1\,\Msol$ and metallicity $Z_0=0.02$ at zero-age.
If this is the case then the first CE phase would have been initiated when star 1 was a first-ascent red giant branch (RGB) star with a degenerate helium core mass $M_{\rmn{core}}$ close to $0.3\,\Msol$.
The effect of this first CE phase is to remove most of the mass of the envelope of star 1, but it is unclear how much of the H-rich envelope remains.
We can use \textsc{mesa/star} to evolve such a star to this stage and find its pre- and post-CE structure.
We prescribe the post-CE structure with a method inspired by that of \citet{1986ApJ...311..742I}: we assume that the composition profile remains constant and the end-of-CE mass is such that the thermal equilibrium post-CE star just fills its Roche lobe \citep{2013MNRAS.435.2048H}.
We can produce such models with the  \textsc{mesa/star} \texttt{relax\_mass} option, provided there is no convective mixing during the artificial mass-loss phase.
With this option the star loses mass at a rate less than about $10^{-3}\,\Msol\,\rmn{yr}^{-1}$ with composition changes from nuclear burning suppressed until its mass reaches a chosen value, $0.3\,\Msol$ in this case.
We choose to remove mass from a star with a $0.28\,\Msol$ core mass, $0.02\,\Msol$ smaller than the chosen post-CE mass.
After the mass-loss phase, the star expands to regain thermal equilibrium and begins to contract after its core mass increases slightly.
The evolution from this point is shown in the effective temperature--luminosity plane (theoretical Hertzsprung--Russell diagram) in Fig.~\ref{fig:Teff-L_HeWD}: the star continues to burn hydrogen in a low-mass shell at roughly constant luminosity, reaches the knee and begins to cool as a WD.
After this stage the star undergoes a H-shell flash which reduces the mass of the envelope by nuclear burning.
We do not include mass-loss by wind or Roche lobe overflow in this phase.
In reality, the star would have a much smaller post-mass-loss radius to fit inside its Roche lobe at the end of the CE phase.
Provided the end-of-CE structure satisfies the chosen prescription then it should be that of the model with the appropriate radius in the contraction sequence; the subsequent evolution to become a WD would then be the same as for our model.

While star 1 evolves in this manner, star 2 remains an MS star.
After several $\rmn{Gyr}$, it too expands to become a red giant, with a core mass just below $0.2\,\Msol$.
At this stage, it fills its Roche lobe and initiates the second CE phase.
We can repeat the mass-loss algorithm used for star 1 to make the CE remnant of star 2, a $0.2\,\Msol$ pre-WD.
When its envelope has been removed by the method described, the star's evolution is similar to that of the $0.3\,\Msol$ star shown in Fig.~\ref{fig:Teff-L_HeWD}, although it does not ignite a shell flash as a WD.
When star 2 detaches from its Roche lobe and reaches the WD cooling track, a short-period detached DWD system is formed.
At the same time, the two WDs move towards one another as the system loses orbital energy and angular momentum through gravitational-wave radiation.
Eventually, WD2 fills its Roche lobe; it is at this point that we wish to know the structure of the two stars.

The fact that both stars may undergo shell flashes means that their precise structure at the start of interaction depends in detail on their separation at the start of the detached phase, the time between this point and when WD2 fills its Roche lobe, and how much mass is lost during the shell flashes.
Further, the precise white dwarf mass range for which shell flashes occur is uncertain; for example, previous modelling has shown that whether flashes occur depends on whether diffusion is included \citetext{\citealp{1999A&A...350...89D}; \citealp*{2001MNRAS.324..617A}}.
Thus, if we are to avoid a full-scale investigation of binary evolution for all zero-age conditions, which would introduce tens of uncertain parameters, it is necessary to make some simplifications in treating the start-of-merger configuration.
One option would be to take both WDs to be at the knee stage, which would give the maximum hydrogen mass at the start of the merger.
This option is likely to overestimate the hydrogen mass for some systems because the WDs continue to burn hydrogen as they cool \citep{1999A&A...350...89D}.
A better approximation would be to allow for this burning that takes place as the WDs spiral towards one another, but to do so requires an estimate of the lifetime as a detached DWD.
We choose $1\,\rmn{Gyr}$ as an estimate of this time between the end of the second mass-loss phase and the onset of interaction.
In the synthetic populations of \citet{2011MNRAS.417.1392Y}, $1\,\rmn{Gyr}$ is roughly the time between the formation of the first He+He DWD and the first He+He DWD merger.
The minimum hydrogen mass at the start of the merger would be if both stars cool sufficiently that they stop burning hydrogen.
We choose $12\,\rmn{Gyr}$ as an estimate of this maximum age of the population.
We record the structure of the WDs at all three stages mentioned: the knee stage (\textbf{young}, the very first point at which they become a WD) and at $1\,\rmn{Gyr}$ (\textbf{middle-aged}) and $12\,\rmn{Gyr}$ (\textbf{old}) after this stage.
The first case is appropriate for a system that merges a short time after the end of the second CE phase; the third case for a system that merges a long time after the end of the second CE phase.
In our example system, the masses of hydrogen in the stars are $M_{\rmn{H,\,WD1}}=1.2 \times 10^{-3}\,\Msol$ and $M_{\rmn{H,\,WD2}} = 2.6 \times 10^{-3}\,\Msol$ if the WDs are young, or $M_{\rmn{H,\,WD1}}= 0.28\times 10^{-3}\,\Msol$ and $M_{\rmn{H,\,WD2}} = 0.62 \times 10^{-3}\,\Msol$ if the WDs are old.

Although we have framed the discussion in terms of a system evolving through a CE+CE channel for a particular pair of zero-age masses, we would expect similar start-of-merger structures for $0.3$+$0.2\,\Msol$ He+He systems that evolved through the RLOF+CE channel or have different zero-age masses.
With our prescription for structures at the end of a CE phase, a post-RLOF system has a similar structure to a post-CE system, but detaches from its Roche lobe at larger orbital separation and with a larger envelope mass: for the same WD mass the stars should have the same envelope mass at the knee stage.
The zero-age mass and pre-mass-loss mass are expected to be less important parameters because red giants have a core structure which mostly depends on the core mass \citep{1968MNRAS.140..387E}.
Thus, we consider the structure of the WDs as discussed to be representative of all $0.3$+$0.2\,\Msol$ He+He mergers.

\subsection{Onset of mass transfer}\label{sec:merger_evolution}
When WD2 first fills its Roche lobe and begins to transfer mass to WD1 there are two possible outcomes: unstable mass transfer and a merger, or stable mass transfer and a semidetached phase.
The predicted outcome is related to the stability of mass transfer at the onset of interaction.
A simple linear treatment of the stability of mass transfer for a cold, degenerate, Roche lobe-filling white dwarf suggests that if the mass ratio $q=M_{\rmn{WD2}}/M_{\rmn{WD1}} \ga 2/3$ then mass transfer is dynamically unstable and a merger begins \citep{1979AcA....29..665T}; otherwise a stable semidetached phase of mass transfer is initiated.
The mass ratio of the example $0.3$+$0.2\,\Msol$ system is on the boundary of instability according to this criterion.
However, predictions for which pairs of WD masses interaction is unstable and leads to a merger become considerably more uncertain on detailed analysis.
Because the mass-transfer rate strongly depends on the degree to which WD2 overfills its Roche lobe, the stability depends on the response of the radius of WD2 and radius of its Roche lobe to mass transfer.
The simple analysis that gives a critical mass ratio $q \approx 2/3$ assumes that mass transfer conserves both mass and orbital angular momentum.
However, \citet{2001A&A...368..939N} and \citet*{2004MNRAS.350..113M} noted that the radius of \emph{WD1} is close to its Roche lobe radius and thus mass transfer can proceed by direct impact rather than disc accretion.
In other words, the accretion stream hits the surface of WD1 directly and thus carries angular momentum from the orbit which adds to the spin of WD1.
The strength of tidal spin--orbit coupling is then important to how quickly this angular momentum is returned to the orbit, how the separation of the WDs changes, and thus the stability of mass transfer.
The most recent sophisticated semi-analytic treatment of the evolution through mass transfer by \citet{2014ApJ...785..157S} and \citet*{2015ApJ...806...76K} built on the earlier work of \citet{2004MNRAS.350..113M} and \citet*{2007ApJ...655.1010G} and considered this uncertain tidal coupling strength, as well as allowing for asynchronous components and the changing size of the critical tidal lobe.
According to the calculations of \citet{2015ApJ...806...76K}, their fig.~6, our example system would undergo a phase of stable mass transfer rather than a merger for a reasonable strong tidal coupling.
In fact some idealized 3D hydrodynamic simulations also agree that an interacting DWD system with $q=2/3$ could avoid merging \citep{2006ApJ...643..381D, 2007ApJ...670.1314M}, while others show that a merger takes place: in a $0.3$+$0.2\,\Msol$ system WD2 is smeared out after a phase of mass transfer perhaps lasting tens of initial orbital periods \citep{2012MNRAS.422.2417D}.
On the other hand, \citet{2015ApJ...805L...6S} points out that the initial phase of accretion in interacting DWDs, even for disc accretion, may lead to merging for all pairs of WD masses.
He considered a process that is not accounted for in the semi-analytic treatments or hydrodynamic simulations: the response of WD1 on accreting H-rich matter from WD2 at the rates expected.
He argued that the initial mass-transfer leads to the formation of a nova-like outburst from the surface of WD1 and initiates a CE phase.
Friction on the WDs inside the CE then causes the separation of the stars to decrease and the onset of a merger.
Mass-transfer stability also depends on the mass--radius relation for WD2, which is affected by the presence of a H-rich envelope \citetext{\citealp{2006ApJ...653.1429D}; \citealp*{2012ApJ...758...64K}; \citealp*{2013ApJ...770L..35S}}.
We assume that WD2 is disrupted in our example system.

\subsection{Mass transfer and disruption phases}
Assuming that mass transfer is unstable, the evolution of the system approaches the dynamical time-scale and a 3D hydrodynamical simulation is a more appropriate description of the ensuing disruption than a 1D stellar-evolution model.
Although many such simulations have been published concerning CO+CO and CO+He DWD mergers \citetext{\citealp*{2009A&A...500.1193L}; \citealp{2012ApJ...757...76S}; \citealp{2013ApJ...770L...8P}; \citealp{2015ApJ...807..105S}; \citealp{2015ApJ...806L...1Z}}, there are few specifically concerning He+He DWD mergers.
Only two groups have studied the dynamical disruption phase for these mergers: \citet*{2004A&A...413..257G} simulated an equal-mass $0.4$+$0.4\,\Msol$ He+He merger and \citet{2011ApJ...737...89D, 2012MNRAS.422.2417D, 2014MNRAS.438...14D} simulated both equal- and unequal-mass He+He mergers.
In fact \citet{2011ApJ...737...89D} made a 3D smoothed particle hydrodynamics simulation relevant to our example: a $0.3$+$0.2\,\Msol$ He+He merger with $4 \times 10^5$ particles in which both WDs are initially synchronized with the $P_0 = 208\,\rmn{s}$ orbit, isothermal with temperatures of $10^5\,\rmn{K}$ and are made of helium only.
They evolve with nuclear burning according to a minimal nuclear reaction network \citep{1998ApJ...503..332H}.
In this simulation, in common with those across the full range of unstable start-of-merger conditions, WD2 is disrupted and within a time span of only tens of $P_0$ a quasi-static remnant is formed.
This remnant can be divided into a central spherically symmetric object (a core and envelope), a disc and a tidal tail.
The core is the roughly isothermal part of the remnant, its outer edge defined as the point at which there is `a sharp increase in temperature', and the inner boundary of the disc is chosen based on the behaviour of angular velocity along the coordinate axes.
For our example, the remnant is made of a core of $0.211\,\Msol$, an envelope of $0.181\,\Msol$, a disc of $0.098\,\Msol$ and a tidal tail of $0.00965\,\Msol$; only $0.00072 \,\Msol$ escapes the system during the disruption phase of the merger.
No significant prompt nucleosynthesis occurs, i.e., the disruption phase is cold and the remnant is composed entirely of helium.
The two WDs are mixed in the disruption phase.
A simple approximation to the simulation data is that the $0.211\,\Msol$ core is the undisturbed core of WD1 and that the other components are a mixture of WD2 and the $0.089\,\Msol$ dredged from the outer layers of WD1. 

\subsection{Post-disruption phase and post-merger phase}
How does the system evolve from this central remnant + disc configuration to become a single star?
There are two major views regarding evolution in the immediate post-disruption phase.
The two views differ in the time-scale on which the matter exterior to the central object is rearranged.
Regardless of which view is correct, eventually the object can be modelled with a 1D stellar-evolution code as a normal evolving star.
The two views imply different structures (mass, composition profile and entropy profile) for the star at the start of this phase, the post-merger phase.
Several methods have been used to find this end-of-merger structure.
These methods normally involve computing the evolution through the post-disruption phase; hence, we discuss the treatment of these two phases together.

\subsubsection{Disc accretion view}
In the disc accretion view, the disc is accreted by the central object.
The first models of the remnants of CO+CO mergers in this view limited the mass accretion rate from this disc to the Eddington rate, the rate for which the Eddington luminosity $L_{\rmn{Edd}} = 4 \pi c GM/\kappa$ equals the accretion luminosity $L_{\rmn{acc}} = G M \dot{M} / R$ \citep{1985ApJ...297..531N}.
This assumption gives a maximum accretion rate of about $10^{-5}\,\Msol\,\rmn{yr}^{-1}$; thus in our example it would take about $10^4\,\rmn{yr}$ for the entire disc to be accreted.
\citet{1990ApJ...353..215I} made the first 1D stellar-evolution computations for He+He DWD mergers in this view.
In the immediate post-disruption phase, a He WD, WD1, accretes He-rich matter at $10^{-5}\,\Msol\,\rmn{yr}^{-1}$ and compressional heating causes a temperature maximum to develop near the core boundary.
In the post-merger phase, when accretion has stopped, helium is ignited in a series of off-centre flashes which progress to the centre and the star becomes a core-He-burning star, a He MS star, a hot subdwarf.
Other cases have been investigated in a similar way by \citet{1998ApJ...500..388S}, \citet{2000MNRAS.313..671S} and \citet{2002MNRAS.336..449H}.
More recently, \citet{2012MNRAS.419..452Z, 2012MNRAS.426L..81Z} have modelled equal-mass mergers in this framework, but, whereas the previous work had treated the central remnant as the undisturbed WD1, these authors first built up a hot envelope by rapid ($10^4\,\Msol\,\rmn{yr}^{-1}$) accretion.
This produces a model with a structure intended to mimic the central remnant: a cold core with a hot envelope.
This central object then accretes from the disc at about $10^{-5}\,\Msol\,\rmn{yr}^{-1}$.
When accretion is complete, the star is a helium giant and, as \citet{1990ApJ...353..215I} found, ignites helium to become a He MS star, although with mixing and burning that depends on the remnant mass.
None of the remnants in this previous work contains hydrogen.

\subsubsection{Viscous evolution view}
According to an alternative view, the differentially rotating matter, the disc, surrounding the central object is expected to have a strong magnetic field.
This means the magnetorotational instability acts, and the resulting large turbulent viscosity leads to heating and rapid redistribution of angular momentum \citep{2012ApJ...748...35S}.
Multi-dimensional simulations of this process by \citet{2012MNRAS.427..190S} show that in about $10^{-4}$--$1\,\rmn{yr}$ the disc is effectively accreted and forms a high-entropy, slowly rotating, solid-body envelope surrounding the central object.
They used the simulations of \citet{2014MNRAS.438...14D} as their initial conditions.
In a simulation of the post-disruption evolution of a $0.3$+$0.2\,\Msol$ He+He case, they found that at the end of the viscous post-disruption phase, the system becomes a He giant.
They made a star with a qualitatively correct entropy profile to make a 1D stellar-evolution model of the post-merger evolution.
It is difficult to follow the composition evolution in the viscous phase so they assumed a He-rich composition.
After a short phase of thermal readjustment, the star ignites helium in a series of off-centre flashes and becomes a core-He-burning star, similar to previous work.
Their remnant did not contain hydrogen.

\subsection{Hydrogen in the He-burning merger remnant}\label{sec:MH_method}
Now that we have discussed the full evolution through a merger to a core-He-burning star, we can discuss the mass of hydrogen in the remnant.
We find that at the start of the $0.3$+$0.2\,\Msol$ merger, there is a total hydrogen mass of $1.6 \times 10^{-3}\,\Msol$ if both WDs are middle-aged.
The simplest assumption is that of a `cold merger', that all of this hydrogen survives the merger and the He-flash phase to give a remnant of total mass $0.5\,\Msol$ made mostly of helium, the pre-merger mass of hydrogen and $Z=0.02$.
However, the discussion in the previous sections has made it clear that there is likely to be some hydrogen burning in the various phases of the merger, although the precise amount remains uncertain and probably depends on whether the accretion or viscous view is more realistic.
We improve on the cold merger assumption by assuming that matter is fully mixed and that any hydrogen that is part of the hot envelope during the disruption phase is likely to be burned at the high temperatures found there.
An alternative assumption to find the mass of hydrogen in the remnant is thus to take the mass of the outer layers of WD1 that are dredged up, according to the simulation of \citet{2014MNRAS.438...14D}, and all the matter from WD2, mix those together and convert to helium the mass of hydrogen that is found in the hot envelope.
We use the formulae by \citet{2014MNRAS.438...14D} to estimate the mass of each component.
The result is that a smaller amount of hydrogen is predicted to remain after the merger of two middle-aged WDs: $0.79 \times 10^{-3}\,\Msol$, about half the pre-merger mass.
This is an estimate of the mass of hydrogen accounting for the fact that some is converted to helium during the merger.
This treatment includes only what we know about the disruption phase of the merger.
We do not know how to estimate how much of this hydrogen may be burned in the post-merger pre-core-He-burning phase without more detailed simulations.
Our estimates provide an extreme to the mass of hydrogen expected to survive to the core-He-burning phase.
Accounting only for this uncertainty, we expect that the true mass of hydrogen is less than this.

\subsection{Evolution in the $T_{\rmn{eff}}$--$\log g$ plane}
Our simple estimates of the mass of hydrogen in the merger remnant can be used, with some extra assumptions, to predict the $T_{\rmn{eff}}$ and $\log g$ of the remnant in the core-He-burning phase.
It is during this phase that the evolution is slowest, and thus in this phase that we expect to find merger remnants.
We use generalized main-sequence (GMS) models to make these predictions.
A GMS is a set of stars with He-burning cores and hydrogen envelopes for which the ratio of core mass to total mass is constant and the total mass varies \citep*{1968ZA.....68..107G}.
A GMS model is then a star of given helium-core and hydrogen envelope mass.
We assume that the merger remnant at the onset of central He burning is a GMS star because the hydrogen has diffused to the surface.
A familiar example of a set of GMS stars is the zero-age horizontal branch, a set of GMS models in which the helium-core mass is fixed at about $0.47\,\Msol$ and the hydrogen envelope mass varies.

We would like GMS stars of different core and envelope masses to agree with realistic ZAEHB stars.
To achieve this, we assume that all GMS stars have a fixed finite transition region, a smoothed step function, from the helium core to the hydrogen envelope over a mass of about $0.3 \times 10^{-3}\,\Msol$ and that the metals are distributed evenly throughout the model.
We take the realistic compositions of the cores and envelopes of the pre-merger WDs.
We assume a solar-like envelope abundance for hydrogen and helium, and a uniform metal abundance with CNO processed metals appropriate to the composition of mixed He WD cores.
These simplified assumptions match the ZAEHB models made by removing mass from low-mass red giants that go on to ignite helium.

\begin{figure}
  \centering
  \includegraphics[width=84mm]{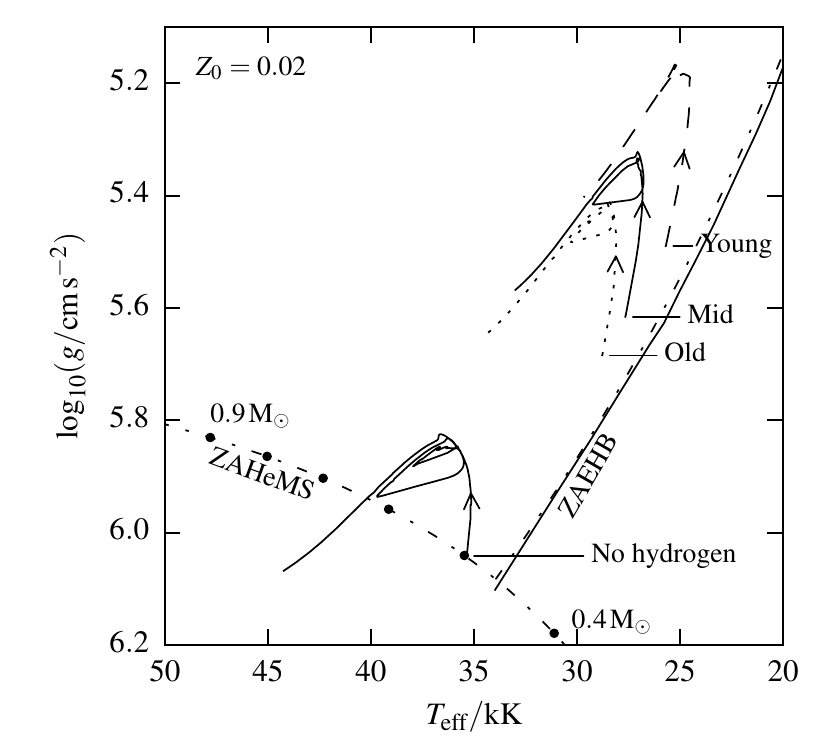}
  \caption{
    Evolution in the effective temperature--gravity plane of $Z_0=0.02$, $M=0.5\,\Msol$ GMS stars and of a naked helium star of the same mass during their core-He-burning phases.
These are possible models of the remnant of a $0.3$+$0.2\,\Msol$ He+He DWD merger in which both pre-merger WDs are at the stages indicated or contain no hydrogen.
A zero-age helium main-sequence (ZAHeMS) is shown with ZAHeMS stars of $0.4$--$0.9\,\Msol$ in steps of $0.1\,\Msol$ indicated by black dots.
Our ZAEHB for a core mass of $0.464\,\Msol$ (dot--dashed line) is shown, as is the ZAEHB computed by (\citealp{1993ApJ...419..596D}, solid line).
}
  \label{fig:Teff-log_g_eg}
\end{figure}

To model the remnant of a $0.3$+$0.2\,\Msol$ merger, we first make a naked helium star with total mass equal to $0.5\,\Msol$ in complete thermal equilibrium ($\epsilon_{\rmn{grav}} = 0$).
We then morph the composition profile of this star to a GMS profile with the appropriate core mass, hydrogen mass and metal mixture, using the \texttt{relax\_initial\_composition} option in \textsc{mesa/star}, which slowly changes the profile of a given star in a series of pseudo-evolutionary steps.
We evolve a naked helium star and a GMS star with the hydrogen mass $0.79 \times 10^{-3}\,\Msol$, the remnant of the merger of middle-aged WDs, to give the evolutionary tracks shown in Fig.~\ref{fig:Teff-log_g_eg}.
We expect that the core-He-burning evolution of the remnant of a $0.3$+$0.2\,\Msol$ He+He DWD merger is somewhere between these two tracks.
We also show the tracks appropriate for pre-merger young and old WDs.
For orientation, we also show a helium MS (a set of GMS stars with zero hydrogen mass and varying total mass) and a ZAEHB for a core mass of $0.464\,\Msol$.
The possible remnants evolve similarly, burning helium until it is exhausted and they become WDs with CO cores and massive helium envelopes.

\subsection{General method}
The method described allows us to estimate the hydrogen mass and positions in the $T_{\rmn{eff}}$--$\log g$ plane during core He burning for other start-of-merger mass pairs.
First, we make a set of seven He WDs with masses between $0.15$ and $0.45\,\Msol$ in steps of $0.05\,\Msol$ by stripping the envelope (using \texttt{relax\_mass}) from a red giant with zero-age mass $M_0=1\,\Msol$ and metallicity $Z_0=0.02$.
We record the masses and compositions of the cores and envelopes at the various ages in their evolution.

The seven white dwarf models allow for $28$ unique mass pairs.
As discussed in Section~\ref{sec:merger_evolution}, it is uncertain whether a merger is expected when WD2 fills its Roche lobe for all these pre-merger WD mass pairs: for some a phase of stable mass transfer is possible.
However, we choose to compute merger remnants for all pairs to allow for the widest possible range of post-merger properties given this uncertainty regarding the stability of mass transfer.
We estimate the post-merger hydrogen mass by the method described in Section~\ref{sec:MH_method}: mass is burned in the hot envelope, the mass of which is computed according the formulae derived from the simulations of \citet{2014MNRAS.438...14D}, which give the mass of the core, inner and outer envelope in each remnant.
Thus, we find the hydrogen mass in the remnant for each pair of pre-merger WD masses and different WD ages, so we can make GMS models with the appropriate total mass and hydrogen mass.

\section{A sample of apparently isolated hot subdwarfs}\label{sec:sample}
Because our interest is in whether DWD merger remnants are a consistent explanation of isolated hot subdwarfs, we need a sample of these stars to compare against our models.
For this purpose, the ideal would be a set of subdwarfs, each with atmospheric parameters measured in a consistent manner and evidence, such as a lack of radial-velocity variations and composite spectra, against the presence of companions.
Even if we had such a sample, it is possible to imagine scenarios in which a subdwarf has a companion hidden beyond detection; for example, a substellar companion on a highly inclined orbit.
Thus, it is very difficult to be certain that any subdwarf is truly isolated.
Although there are several published samples of hot subdwarfs, few satisfy our criteria.
Perhaps the best available is the sample of \citet{2012A&A...543A.149G}.
They gave the designations of some single-lined systems which do not show evidence of radial-velocity variability.
Unfortunately, their table 1 does not provide the $T_{\rmn{eff}}$ and $\log g$ for these systems, although a subset do have these parameters recorded in a later paper \citep{2013A&A...557A.122G}.
Thus we start with this set of systems, summarized in their table 2.
We compare this list of systems with the earlier work \citep{2012A&A...543A.149G} and remove those with double-lined spectra or variable radial velocities to give $37$ stars.
This is a set of systems with single-lined spectra and evidence against radial-velocity variability.
Errors on the parameters for each individual system were not given by the authors, but they did give representative values of $\Delta T_{\rmn{eff}}= 1.1\,\rmn{kK}$ and $\Delta \log g =0.12$.
Evidence for a lack of radial-velocity variations in EC\,$22081{-}1916$ was reported by \citet*{2011ApJ...733L..13G}.
We add to our sample the fast-rotating sdB SB\,$290$ \citep{2013A&A...551L...4G}.
We also add the $10$ stars identified as single with only one set of atmospheric parameters in table 1 of \citet{2012A&A...539A..12F}.
Our full sample of $48$ sdBs is summarized in Table~\ref{tab:sample}.

\section{Results}\label{sec:results}
We first describe our He WD models.
We then give the hydrogen mass in the merger remnants.
These results are used to find the region in the $T_{\rmn{eff}}$--$\log g$ where He+He DWD merger remnants are likely to be found.
We check whether the observed apparently isolated hot subdwarfs lie in this region.

\subsection{Full set of He white dwarfs}
\begin{figure}
  \centering
  \includegraphics[width=84mm]{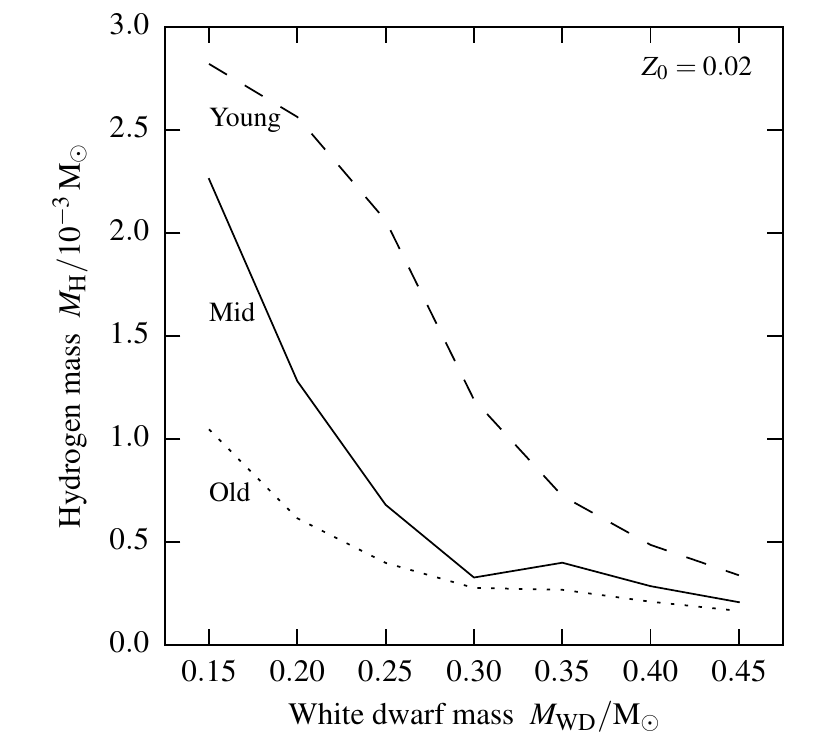}
  \caption{
Mass of hydrogen in He WDs, made by removing mass from a $1\,\Msol$ star, of masses $M_{\rmn{WD}}$ and $Z_0=0.02$.
The mass of hydrogen is given for three stages: at the knee, a young WD; at $1\,\rmn{Gyr}$ after the knee, a middle-aged WD; and at $12\,\rmn{Gyr}$ after the knee, an old WD.}
  \label{fig:M_WD-M_H1}
\end{figure}

We used the method discussed in Section~\ref{sec:pre_merger_evolution} to make seven He WDs.
Of these, only the $0.25$ and $0.3\,\Msol$ WDs undergo shell flashes.
Fig.~\ref{fig:M_WD-M_H1} shows the mass of hydrogen as a function of WD mass in these stars at three different stages: the knee stage, when they first reach the WD cooling track (young); $1\,\rmn{Gyr}$ after the knee stage (middle-aged); and $12\,\rmn{Gyr}$ after the knee stage (old).
Generally, the hydrogen mass decreases with increasing WD mass.
There is a small dip in the curve for middle-aged (mid) WDs attributable to the occurrence of H-shell flashes.

There is a complication in computing the post-CE evolution of stars with core masses less than about $0.25\,\Msol$ that could affect the hydrogen mass in their WD progeny.
Such stars have less sharply defined core--envelope structure than RGB stars with more massive cores.
Mass has to be removed into H-deficient layers if the post-mass-loss star is to contract.
It is possible that the mass-loss method and post-mass-loss evolution affect the evolution, so that there is not a unique relation between the two.
We have compared our results to previous work that considered post-RLOF WDs and found hydrogen masses for the lower mass WDs considered here that are consistent with such work \citep{1998A&A...339..123D}.

\subsection{Hydrogen mass in merger remnants}
\begin{figure}
  \centering
  \includegraphics[width=84mm]{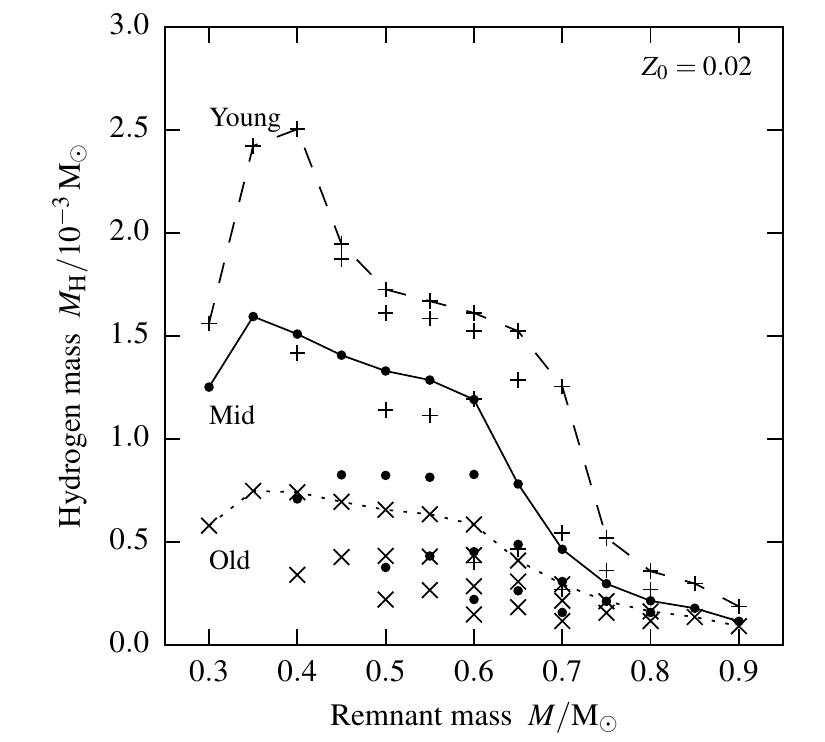}
  \caption{
Total mass of hydrogen in He+He DWD merger remnants for given remnant mass $M$ and $Z_0=0.02$.
There are multiple hydrogen masses for each remnant mass because more than one combination of pre-merger WD masses can produce the same remnant mass.
Crosses represent old pre-merger WDs and the dotted line is the maximum for a given remnant mass.
Circles represent middle-aged pre-merger WDs and the solid line is the maximum for a given remnant mass.
Pluses represent young pre-merger WDs and the dashed line is the maximum for a given remnant mass.}
  \label{fig:remnant_mass-M_H1}
\end{figure}

Fig.~\ref{fig:remnant_mass-M_H1} shows the hydrogen mass for all unique combinations of the WDs computed at the young, middle-aged and old stages.
There is a typical hydrogen mass in a remnant of up to about $1.6 \times 10^{-3}\,\Msol$ for pre-merger middle-aged WDs.
\citet{2002MNRAS.336..449H} assumed an envelope mass of $10^{-3}\,\Msol$.
If their model envelopes had a composition of about $X=0.7$, the maximum hydrogen mass was about $0.7 \times 10^{-3}\,\Msol$, smaller than the maximum we find for pre-merger middle-aged WDs.

\subsection{Comparison with observed hot subdwarfs in the $T_{\rmn{eff}}$--$\log g$ plane}
\begin{figure}
  \centering
  \includegraphics[width=84mm]{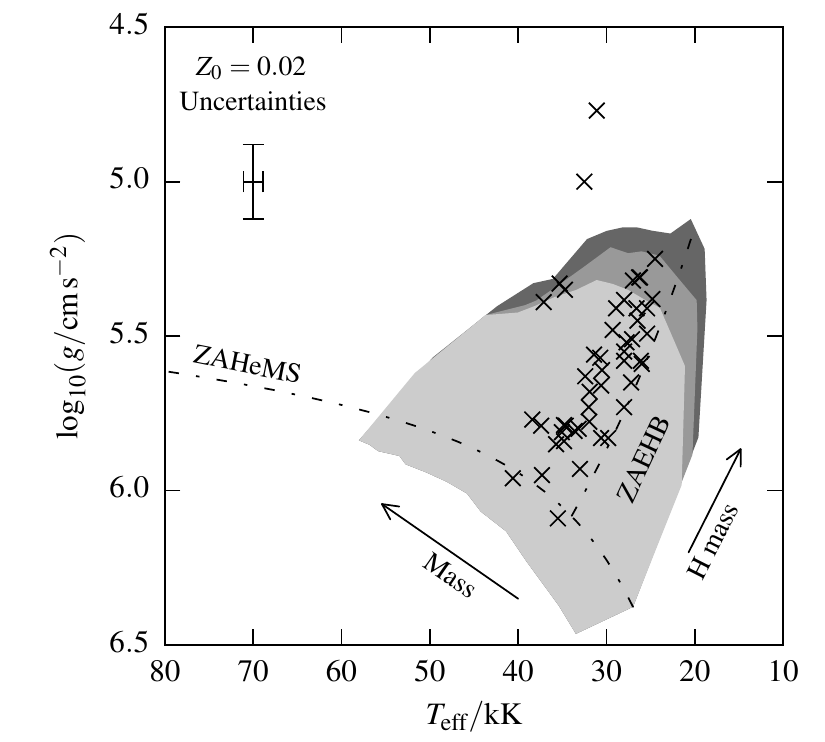}
  \caption{
Effective temperature--log surface gravity plane.
Sample of observed apparently isolated hot subdwarfs (Table~\ref{tab:sample}).
The shaded regions give the position of $Z_0=0.02$ He+He DWD merger remnants during core He burning (when the central helium abundance exceeds $10^{-3}$) for different assumptions about the age of the pre-merger WDs.
The region for old pre-merger WDs is shaded in the lightest grey.
The region for middle-aged pre-merger WDs is the union of this region and the next darkest grey.
The region for young pre-merger WDs is the entire shaded region.
The arrows roughly indicate the directions of increasing total mass and hydrogen mass at the start of central He burning.
The ranges of these two quantities are as in Fig.~\ref{fig:remnant_mass-M_H1}.
}
  \label{fig:Teff-log_g_regions}
\end{figure}

Are our model merger remnants consistent with apparently isolated He-poor hot subdwarfs?
We compute GMS stars for the remnant mass--maximum hydrogen mass combinations appropriate to each of the three possible pre-merger WD stages.
Because we have neglected mass-loss from the WDs and there may be more hydrogen burning in the post-merger phase than estimated, we also compute naked helium star models for each remnant mass.
These models are used to find the region occupied by core He-burning models, when the central helium mass fraction exceeds $10^{-3}$.
Fig.~\ref{fig:Teff-log_g_regions} shows the stars from our sample in the $T_{\rmn{eff}}$--$\log g$ plane and the shaded region allowed by the evolution of our GMS stars as remnants of mergers of WDs of different ages.
A He+He DWD merger of mass $0.15$+$0.15\,\Msol$ produces a $0.3\,\Msol$ remnant, a mass which is too low for a core-He-burning star.
Typically, the minimum mass for the ignition of helium core is about $0.33\,\Msol$, but if the star is the remnant of a DWD merger, \citet{2002MNRAS.336..449H} showed that the minimum mass can depend on the details of accretion.
We take $0.33\,\Msol$ as the minimum remnant mass and interpolate to find the hydrogen mass for such a remnant mass.
The formulae given by \citet{2014MNRAS.438...14D} only considered pre-merger WD masses greater than $0.2\,\Msol$.
Fortunately, their formulae are given in terms of total mass $M$ and mass ratio $q$, so it is straightforward to extrapolate to mergers involving $0.15\,\Msol$ WDs.

Fig.~\ref{fig:Teff-log_g_regions} shows that it is quite possible to explain the majority of our sample as post-DWD merger objects, irrespective of the age of the pre-merger WDs.
Even those stars that have surface gravities slightly too low to fit in the old pre-merger WD region are consistent within the typical errors.
Only the two stars with particularly low surface gravities, EC\,$22081{-}1916$ and EC\,$20229{-}3716$, are difficult to explain as He+He DWD merger remnants for any age of the pre-merger WDs.
These conclusions require that all observed stars are in the core-He-burning phase.

There are several uncertainties that may change the computed $T_{\rmn{eff}}$--$\log g$ regions and thus affect the degree to which our models explain the sample.

\subsubsection{Pre-merger configuration}
In our treatment of the pre-merger evolution we have generally made the assumptions most favourable to the retention of hydrogen.
This is particularly the case for our assumptions about the structure of CE remnants.
For example, the alternative prescription of \citet{2011ApJ...730...76I}, with which much lower mass envelopes remain after the end of CE phases \citep{2013MNRAS.435.2048H}, would give lower hydrogen masses in the pre-merger WDs and make the regions lie closer to the ZAHeMS.
Another way in which we may have overestimated the hydrogen mass in the He WDs, and thus in the merger remnants, is by neglecting diffusion.
The range of WD masses for which shell flashes occur is sensitive to diffusion, as is the mass of hydrogen that is burned in these flashes \citetext{\citealp{1986ApJ...311..742I}; \citealp*{2001MNRAS.323..471A}}.
Models that include diffusion and undergo shell flashes burn more hydrogen and so that models including pre-merger WDs with diffusion may move the regions closer to the ZAHeMS.
Observational constraints on the envelope masses of He-core WDs would help to reduce these uncertainties.
Particularly, study of pulsating low-mass WDs \citep{2012ApJ...750L..28H} and pulsating pre-WDs \citep{2013Natur.498..463M} will constrain the envelope mass of these stars \citep{2012A&A...547A..96C}.
The DWD binary system CSS\,$41177$ \citep{2011ApJ...735L..30P, 2014MNRAS.438.3399B} could also be used to constrain the envelope masses of the two low-mass WDs because it is a double-lined eclipsing system.

\subsubsection{Merger treatment}
A crucial assumption to the validity of our approach, particularly with respect to the mass of hydrogen which survives the merger, depends on WD2, the lighter WD, being disrupted and mixed in the merger.
Generally, the start-of-merger configurations in merger simulations are necessarily approximations to the more realistic structures discussed earlier.
In particular, the envelopes of the two white dwarfs have often been neglected.
While there are no simulations of mergers of He WDs with H-rich envelopes, there are simulations of mergers of COWDs with H- or helium-rich envelopes \citep{2012ApJ...746...62R, 2013ApJ...770L...8P}.
These CO+CO simulations could give some hints as to what He+He simulations with envelopes would look like.
In the CO+CO case, the high temperature in the shock formed in direct impact leads the outer layers to be simply burned immediately: very little of the envelope survives the disruption phase of the merger.
If this is the case for He WDs as well, then it is reasonable to assume that hydrogen is not present in He+He merger remnants.
However, the early stages of mass transfer in the merger are not well resolved (the simulations start with very large mass-transfer rates) and do not include physics such as radiative transfer.
Including radiation may move the outer layers to large radius as a CE forms in response to super-Eddington mass-transfer rates in the early phases of the merger \citep{1988ApJ...324..355I}.
The hydrogen could then become mixed in the envelope and disc.
An observation which could be relevant to this discussion is that some R Coronae Borealis stars, some of which seem consistent with a CO+He DWD merger origin, show traces of hydrogen in their spectra \citep{1994JApA...15...47L}.
If it is reasonable to assume that hydrogen survives the disruption of a He WD in that case, then it is reasonable to assume the same for a He+He DWD merger.
Future simulations may check our assumption about the distribution of hydrogen in the merger remnant.

\section{Discussion}\label{sec:discussion}
Our results are most relevant to understanding the outcome of He+He DWD mergers and the formation of isolated hot subdwarfs.
By considering realistic initial conditions in our models, we have found the regions in which merger remnants are likely to be found (with H-rich surfaces) in the long-lived core-He-burning phase.
The majority of the apparently isolated hot subdwarfs we have identified are consistent with having formed through the He+He DWD merger channel on the basis of their $T_{\rmn{eff}}$ and $\log g$.
Only two systems do not fit into any of the $T_{\rmn{eff}}$--$\log g$ regions we have computed: EC\,$22081{-}1916$ and EC\,$20229{-}3716$.
\citet{2011ApJ...733L..13G} had already noted that EC\,$22081{-}1916$, a rapidly rotating low-gravity sdB, has atmospheric parameters that imply a particularly massive H-rich envelope, at odds with its being a He+He DWD merger remnant.
Our calculations confirm this conclusion.

\subsection{Other channels}
He+He DWD mergers are not the only proposed channel to forming isolated hot subdwarfs.
It is thus interesting to ask whether the other proposed channels can account for the hydrogen masses and the $T_{\rmn{eff}}$--$\log g$ regions derived here for isolated hot subdwarfs; can we distinguish between different formation channels on this basis?

One proposed channel is enhanced mass-loss from a single star \citep{1993ApJ...407..649C,1996ApJ...466..359D}, perhaps because of He mixing driven by internal rotation \citep{1997ApJ...474L..23S} or because of an instability associated with positive binding energy \citep*{1994MNRAS.270..121H}.
The key proposal is that there is a sufficiently wide distribution of mass-loss rates for isolated stars of a given zero-age mass and metallicity that there is a correspondingly wide distribution in envelope mass when the stars ignite helium.
Depending precisely on what this distribution is, there should be no difficulty producing a wide range of hydrogen envelope masses and explaining all of the stars in our sample.

Mergers of He WDs with hybrid CO(He) WDs (WDs with massive, $\approx 0.1\,\Msol$ helium envelopes) have also been proposed to produce hot subdwarfs of various types \citep*{2011MNRAS.410..984J}.
The merger process, the disruption of a He WD, should be similar to the He+He DWD mergers considered in this work.
We would thus expect hydrogen masses of a similar order, or perhaps slightly lower than calculated here, because the hybrid CO(He) WDs may have lower hydrogen envelope masses.

Mergers of He WDs with low-mass MS stars have also been proposed to produce hot subdwarfs \citep{2011ApJ...733L..42C}.
Because of the larger mass of hydrogen available in such mergers we expect these to be able to produce a larger range of hydrogen envelope masses in the core-He-burning phase than the He+He DWD mergers considered in this paper.

Mergers of MS stars with RGB stars in a CE, with subsequent enhanced mass-loss attributable to fast rotation, have also been proposed to produce hot subdwarfs \citep{2008ApJ...687L..99P}.
We expect these to also produce a wide range of hydrogen masses for the same reason as in HeWD+MS mergers.

In summary, it seems possible that some of the other channels to the formation of isolated hot subdwarfs could explain the stars in our sample, including the low-gravity outliers.
Another possibility for explaining these low-gravity subdwarfs could be simply that they have an undetected companion in a long-period orbit, or are in a pre-core-He-burning phase.
The uncertainties mean that without further calculation it will be difficult to distinguish the various proposed channels on the basis of their hydrogen masses or positions in the $T_{\rmn{eff}}$--$\log g$ plane.
We have shown that the majority of the apparently isolated hot subdwarfs in our sample can be explained as the remnants of He+He DWD mergers.

\subsection{H- or He-rich hot subdwarfs}
Although our main concern has been the atmospheric parameters $T_{\rmn{eff}}$ and $\log g$, the composition of the surface is clearly essential in understanding the evolution of hot subdwarfs.
One question is whether H-rich (sdO, sdB) or He-rich (He-sdB, He-sdO) subdwarfs are produced by He+He DWD mergers.
In this paper we have argued, in common with \citet{2002MNRAS.336..449H,2003MNRAS.341..669H}, that H-rich hot subdwarfs can be produced by He+He DWD mergers.
Other authors have argued that He+He DWD mergers make He-rich hot subdwarfs because hydrogen is burned in the merger.
For example, \citet{2012MNRAS.419..452Z} considered He+He DWD mergers to form He-sdBs and He-sdOs.
\citet{2011MNRAS.410..984J} suggested that all surface types are possible from DWD merger remnants.
They suggested that the effective temperature of the merger remnant dictates whether the atmosphere is convective and thus whether diffusion can work to produce a He-poor surface from a mixture containing hydrogen.
In their scheme, hotter remnants have He-rich atmospheres while cooler remnants have H-rich atmospheres.
They proposed that He+He DWD mergers produce sdBs, He-sdBs and some He-sdOs, while CO(He)+He DWD mergers produce He-sdOs, because their final effective temperatures are higher.
They suggested that some He+He DWD mergers may make one composition subclass of He-sdO and hybrid+He DWD mergers make another subclass of He-sdO.
These composition subclasses were explained by \citet{2012MNRAS.419..452Z} as the result of mergers of different total mass.

The models presented here require that the diffusion time-scale is sufficiently short that hydrogen diffuses to the surface by or soon after the start of central helium burning.
Time-dependent diffusion calculations show that the diffusion time-scale is short compared to the lifetime of core He burning for typical sdB stars \citep{2010A&A...511A..87H}.
However, in such calculations, diffusion alone is too efficient to explain the observed helium abundances.
This is just one general unexplained observation regarding the detailed surface compositions that our models do not address.
Another is that the hot subdwarfs in Table~\ref{tab:sample} show a range of helium-to-hydrogen number ratios; in fact, \citet{2003A&A...400..939E} found that helium abundance increases with effective temperature along two sequences.
Our models all have solar helium-to-hydrogen number ratio.
The details of mixing and separation in the outer layers of hot subdwarfs are still to be understood \citep{2016PASP..128h2001H}.

\subsection{Mass distribution}
Our results could be interpreted as evidence against the H-rich hot subdwarfs being He+He DWD merger remnants, in the sense that only a small part of our computed region, corresponding to a narrow range in mass, is populated by the observed stars.
\citet{2012A&A...539A..12F} came to a similar conclusion, on the basis of asteroseismology: the observed mass range is narrower than predicted if the merger channel is significant.
However, it should be noted that our work gives only the region expected for He+He DWD merger remnants, not the expected distribution within this region;
the mass distribution in the model populations of \citet{2003MNRAS.341..669H} would also populate a wider region in this plane than occupied by our sample of isolated hot subdwarfs.
At the same time, we are unsure whether this sample of isolated hot subdwarfs is representative of the population as a whole.
These two concerns mean that the implications of only a narrow part of the region being populated are unclear.

\subsection{Surface rotation}
Another potential constraint on the formation of isolated hot subdwarfs is their surface rotation.
The naive prediction is that merger remnants should be fast rotators because of the large orbital angular momentum at the start of a merger.
The relatively slow rotation of the surface of isolated hot subdwarfs has thus been interpreted as evidence against their being merger remnants \citep{2012A&A...543A.149G}.
Of our sample, only SB~$290$ and EC\,$22081{-}1916$ have high surface rotation rates (Table~\ref{tab:sample}).
SB~$290$ can be explained as a core He-burning merger remnant on the basis of its $T_{\rmn{eff}}$ and $\log g$, but EC\,$22081{-}1916$ cannot.
It is unclear whether slow rotation should be taken as evidence against post-merger status, because it is possible that angular momentum is lost by shedding mass in the pre-core-He-burning phase \citep{2012A&A...543A.149G}.
Such mass-loss could reduce the mass of hydrogen in the core He-burning remnant compared to the calculations we have given here.

\section{Conclusion}\label{sec:conclusion}
We have estimated the mass of hydrogen that is present in the remnants of He+He DWD mergers and thus found the region occupied by these stars in the effective temperature--surface gravity plane during the core-He-burning phase.
By comparing to a sample of apparently isolated hot subdwarfs, we find that the majority of these stars can be explained as the remnants of He+He DWD mergers on this basis.
We have identified several uncertainties that could either increase or decrease the extent of this region and change this conclusion.
Reduction of these uncertainties will come through consideration of the following key questions.
What is the mass of hydrogen in He WDs in the start-of-merger configuration?
Asteroseismology of low-mass WDs and pre-WDs could constrain the envelope mass of these stars and check the models computed here.
How much hydrogen survives the merger?
Because hydrogen is at the surface of the WDs, the answer depends crucially on the evolution in the early stages of the merger.
We hope that future work can check our assumption that hydrogen is distributed through the hot envelope and disc in the remnant; such mixing could occur by the formation of a CE in the early, currently numerically unresolved, stages of mergers.
What happens to hydrogen in the post-merger phase?
Which of the disc accretion or viscous view is the more appropriate description?
An answer to one of the main outstanding questions on hot subdwarfs -- how are the isolated examples formed? -- would be greatly helped by a sample of such stars that have been checked for companions across a wide range of parameter space.

\section*{Acknowledgements}
The Armagh Observatory is supported by a grant from the Northern Ireland Department of Culture, Arts and Leisure.
This work was supported by grant ST/M000834/1 from the UK Science and Technology Facilities Council, STFC.

\bibliographystyle{mnras}
\bibliography{paper}

\begin{thebibliography}{}
\makeatletter
\relax
\def\mn@urlcharsother{\let\do\@makeother \do\$\do\&\do\#\do\^\do\_\do\%\do\~}
\def\mn@doi{\begingroup\mn@urlcharsother \@ifnextchar [ {\mn@doi@}
  {\mn@doi@[]}}
\def\mn@doi@[#1]#2{\def\@tempa{#1}\ifx\@tempa\@empty \href
  {http://dx.doi.org/#2} {doi:#2}\else \href {http://dx.doi.org/#2} {#1}\fi
  \endgroup}
\def\mn@eprint#1#2{\mn@eprint@#1:#2::\@nil}
\def\mn@eprint@arXiv#1{\href {http://arxiv.org/abs/#1} {{\tt arXiv:#1}}}
\def\mn@eprint@dblp#1{\href {http://dblp.uni-trier.de/rec/bibtex/#1.xml}
  {dblp:#1}}
\def\mn@eprint@#1:#2:#3:#4\@nil{\def\@tempa {#1}\def\@tempb {#2}\def\@tempc
  {#3}\ifx \@tempc \@empty \let \@tempc \@tempb \let \@tempb \@tempa \fi \ifx
  \@tempb \@empty \def\@tempb {arXiv}\fi \@ifundefined
  {mn@eprint@\@tempb}{\@tempb:\@tempc}{\expandafter \expandafter \csname
  mn@eprint@\@tempb\endcsname \expandafter{\@tempc}}}

\bibitem[\protect\citeauthoryear{{Alastuey} \& {Jancovici}}{{Alastuey} \&
  {Jancovici}}{1978}]{1978ApJ...226.1034A}
{Alastuey} A.,  {Jancovici} B.,  1978, \mn@doi [\apj] {10.1086/156681}, \href
  {http://adsabs.harvard.edu/abs/1978ApJ...226.1034A} {226, 1034}

\bibitem[\protect\citeauthoryear{{Althaus}, {Serenelli}  \&
  {Benvenuto}}{{Althaus} et~al.}{2001a}]{2001MNRAS.323..471A}
{Althaus} L.~G.,  {Serenelli} A.~M.,   {Benvenuto} O.~G.,  2001a, \mn@doi
  [\mnras] {10.1046/j.1365-8711.2001.04227.x}, \href
  {http://adsabs.harvard.edu/abs/2001MNRAS.323..471A} {323, 471}

\bibitem[\protect\citeauthoryear{{Althaus}, {Serenelli}  \&
  {Benvenuto}}{{Althaus} et~al.}{2001b}]{2001MNRAS.324..617A}
{Althaus} L.~G.,  {Serenelli} A.~M.,   {Benvenuto} O.~G.,  2001b, \mn@doi
  [\mnras] {10.1046/j.1365-8711.2001.04324.x}, \href
  {http://adsabs.harvard.edu/abs/2001MNRAS.324..617A} {324, 617}

\bibitem[\protect\citeauthoryear{{Angulo} et~al.,}{{Angulo}
  et~al.}{1999}]{1999NuPhA.656....3A}
{Angulo} C.  et~al., 1999, \mn@doi [\nphysa] {10.1016/S0375-9474(99)00030-5},
  \href {http://adsabs.harvard.edu/abs/1999NuPhA.656....3A} {656, 3}

\bibitem[\protect\citeauthoryear{{Bours} et~al.,}{{Bours}
  et~al.}{2014}]{2014MNRAS.438.3399B}
{Bours} M.~C.~P.  et~al., 2014, \mn@doi [\mnras] {10.1093/mnras/stt2453}, \href
  {http://adsabs.harvard.edu/abs/2014MNRAS.438.3399B} {438, 3399}

\bibitem[\protect\citeauthoryear{{Brown}}{{Brown}}{1995}]{1995ApJ...440..270B}
{Brown} G.~E.,  1995, \mn@doi [\apj] {10.1086/175268}, \href
  {http://adsabs.harvard.edu/abs/1995ApJ...440..270B} {440, 270}

\bibitem[\protect\citeauthoryear{{Buchler} \& {Yueh}}{{Buchler} \&
  {Yueh}}{1976}]{1976ApJ...210..440B}
{Buchler} J.~R.,  {Yueh} W.~R.,  1976, \mn@doi [\apj] {10.1086/154847}, \href
  {http://adsabs.harvard.edu/abs/1976ApJ...210..440B} {210, 440}

\bibitem[\protect\citeauthoryear{{Cassisi}, {Potekhin}, {Pietrinferni},
  {Catelan}  \& {Salaris}}{{Cassisi} et~al.}{2007}]{2007ApJ...661.1094C}
{Cassisi} S.,  {Potekhin} A.~Y.,  {Pietrinferni} A.,  {Catelan} M.,   {Salaris}
  M.,  2007, \mn@doi [\apj] {10.1086/516819}, \href
  {http://adsabs.harvard.edu/abs/2007ApJ...661.1094C} {661, 1094}

\bibitem[\protect\citeauthoryear{{Castellani} \& {Castellani}}{{Castellani} \&
  {Castellani}}{1993}]{1993ApJ...407..649C}
{Castellani} M.,  {Castellani} V.,  1993, \mn@doi [\apj] {10.1086/172547},
  \href {http://adsabs.harvard.edu/abs/1993ApJ...407..649C} {407, 649}

\bibitem[\protect\citeauthoryear{{Caughlan} \& {Fowler}}{{Caughlan} \&
  {Fowler}}{1988}]{1988ADNDT..40..283C}
{Caughlan} G.~R.,  {Fowler} W.~A.,  1988, \mn@doi [At.\ Data Nucl.\ Data
  Tables] {10.1016/0092-640X(88)90009-5}, \href
  {http://adsabs.harvard.edu/abs/1988ADNDT..40..283C} {40, 283}

\bibitem[\protect\citeauthoryear{{Clausen} \& {Wade}}{{Clausen} \&
  {Wade}}{2011}]{2011ApJ...733L..42C}
{Clausen} D.,  {Wade} R.~A.,  2011, \mn@doi [\apjl]
  {10.1088/2041-8205/733/2/L42}, \href
  {http://adsabs.harvard.edu/abs/2011ApJ...733L..42C} {733, L42}

\bibitem[\protect\citeauthoryear{{C{\'o}rsico}, {Romero}, {Althaus}  \&
  {Hermes}}{{C{\'o}rsico} et~al.}{2012}]{2012A&A...547A..96C}
{C{\'o}rsico} A.~H.,  {Romero} A.~D.,  {Althaus} L.~G.,   {Hermes} J.~J.,
  2012, \mn@doi [\aap] {10.1051/0004-6361/201220114}, \href
  {http://adsabs.harvard.edu/abs/2012A%26A...547A..96C} {547, A96}

\bibitem[\protect\citeauthoryear{{Cox} \& {Giuli}}{{Cox} \&
  {Giuli}}{1968}]{1968pss..book.....C}
{Cox} J.~P.,  {Giuli} R.~T.,  1968, Principles of Stellar Structure.
{Gordon and Breach}, {New York}

\bibitem[\protect\citeauthoryear{{Cox} \& {Salpeter}}{{Cox} \&
  {Salpeter}}{1961}]{1961ApJ...133..764C}
{Cox} J.~P.,  {Salpeter} E.~E.,  1961, \mn@doi [\apj] {10.1086/147082}, \href
  {http://adsabs.harvard.edu/abs/1961ApJ...133..764C} {133, 764}

\bibitem[\protect\citeauthoryear{{D'Antona}, {Ventura}, {Burderi}  \&
  {Teodorescu}}{{D'Antona} et~al.}{2006}]{2006ApJ...653.1429D}
{D'Antona} F.,  {Ventura} P.,  {Burderi} L.,   {Teodorescu} A.,  2006, \mn@doi
  [\apj] {10.1086/507408}, \href
  {http://adsabs.harvard.edu/abs/2006ApJ...653.1429D} {653, 1429}

\bibitem[\protect\citeauthoryear{{D'Cruz}, {Dorman}, {Rood}  \&
  {O'Connell}}{{D'Cruz} et~al.}{1996}]{1996ApJ...466..359D}
{D'Cruz} N.~L.,  {Dorman} B.,  {Rood} R.~T.,   {O'Connell} R.~W.,  1996,
  \mn@doi [\apj] {10.1086/177515}, \href
  {http://adsabs.harvard.edu/abs/1996ApJ...466..359D} {466, 359}

\bibitem[\protect\citeauthoryear{{D'Souza}, {Motl}, {Tohline}  \&
  {Frank}}{{D'Souza} et~al.}{2006}]{2006ApJ...643..381D}
{D'Souza} M.~C.~R.,  {Motl} P.~M.,  {Tohline} J.~E.,   {Frank} J.,  2006,
  \mn@doi [\apj] {10.1086/500384}, \href
  {http://adsabs.harvard.edu/abs/2006ApJ...643..381D} {643, 381}

\bibitem[\protect\citeauthoryear{{Dan}, {Rosswog}, {Guillochon}  \&
  {Ramirez-Ruiz}}{{Dan} et~al.}{2011}]{2011ApJ...737...89D}
{Dan} M.,  {Rosswog} S.,  {Guillochon} J.,   {Ramirez-Ruiz} E.,  2011, \mn@doi
  [\apj] {10.1088/0004-637X/737/2/89}, \href
  {http://adsabs.harvard.edu/abs/2011ApJ...737...89D} {737, 89}

\bibitem[\protect\citeauthoryear{{Dan}, {Rosswog}, {Guillochon}  \&
  {Ramirez-Ruiz}}{{Dan} et~al.}{2012}]{2012MNRAS.422.2417D}
{Dan} M.,  {Rosswog} S.,  {Guillochon} J.,   {Ramirez-Ruiz} E.,  2012, \mn@doi
  [\mnras] {10.1111/j.1365-2966.2012.20794.x}, \href
  {http://adsabs.harvard.edu/abs/2012MNRAS.422.2417D} {422, 2417}

\bibitem[\protect\citeauthoryear{{Dan}, {Rosswog}, {Br{\"u}ggen}  \&
  {Podsiadlowski}}{{Dan} et~al.}{2014}]{2014MNRAS.438...14D}
{Dan} M.,  {Rosswog} S.,  {Br{\"u}ggen} M.,   {Podsiadlowski} P.,  2014,
  \mn@doi [\mnras] {10.1093/mnras/stt1766}, \href
  {http://adsabs.harvard.edu/abs/2014MNRAS.438...14D} {438, 14}

\bibitem[\protect\citeauthoryear{{Deca} et~al.,}{{Deca}
  et~al.}{2012}]{2012MNRAS.421.2798D}
{Deca} J.  et~al., 2012, \mn@doi [\mnras] {10.1111/j.1365-2966.2012.20483.x},
  \href {http://adsabs.harvard.edu/abs/2012MNRAS.421.2798D} {421, 2798}

\bibitem[\protect\citeauthoryear{{Dorman}, {Rood}  \& {O'Connell}}{{Dorman}
  et~al.}{1993}]{1993ApJ...419..596D}
{Dorman} B.,  {Rood} R.~T.,   {O'Connell} R.~W.,  1993, \mn@doi [\apj]
  {10.1086/173511}, \href {http://adsabs.harvard.edu/abs/1993ApJ...419..596D}
  {419, 596}

\bibitem[\protect\citeauthoryear{{Driebe}, {Schoenberner}, {Bloecker}  \&
  {Herwig}}{{Driebe} et~al.}{1998}]{1998A&A...339..123D}
{Driebe} T.,  {Schoenberner} D.,  {Bloecker} T.,   {Herwig} F.,  1998, \aap,
  \href {http://adsabs.harvard.edu/abs/1998A%26A...339..123D} {339, 123}

\bibitem[\protect\citeauthoryear{{Driebe}, {Bl{\"o}cker}, {Sch{\"o}nberner}  \&
  {Herwig}}{{Driebe} et~al.}{1999}]{1999A&A...350...89D}
{Driebe} T.,  {Bl{\"o}cker} T.,  {Sch{\"o}nberner} D.,   {Herwig} F.,  1999,
  \aap, \href {http://adsabs.harvard.edu/abs/1999A%26A...350...89D} {350, 89}

\bibitem[\protect\citeauthoryear{{Drilling}, {Jeffery}, {Heber}, {Moehler}  \&
  {Napiwotzki}}{{Drilling} et~al.}{2013}]{2013A&A...551A..31D}
{Drilling} J.~S.,  {Jeffery} C.~S.,  {Heber} U.,  {Moehler} S.,   {Napiwotzki}
  R.,  2013, \mn@doi [\aap] {10.1051/0004-6361/201219433}, \href
  {http://adsabs.harvard.edu/abs/2013A%26A...551A..31D} {551, A31}

\bibitem[\protect\citeauthoryear{{Edelmann}, {Heber}, {Hagen}, {Lemke},
  {Dreizler}, {Napiwotzki}  \& {Engels}}{{Edelmann}
  et~al.}{2003}]{2003A&A...400..939E}
{Edelmann} H.,  {Heber} U.,  {Hagen} H.-J.,  {Lemke} M.,  {Dreizler} S.,
  {Napiwotzki} R.,   {Engels} D.,  2003, \mn@doi [\aap]
  {10.1051/0004-6361:20030135}, \href
  {http://adsabs.harvard.edu/abs/2003A%26A...400..939E} {400, 939}

\bibitem[\protect\citeauthoryear{{Eggleton}}{{Eggleton}}{1968}]{1968MNRAS.140..387E}
{Eggleton} P.~P.,  1968, \mnras, \href
  {http://adsabs.harvard.edu/abs/1968MNRAS.140..387E} {140, 387}

\bibitem[\protect\citeauthoryear{{Ferguson}, {Alexander}, {Allard}, {Barman},
  {Bodnarik}, {Hauschildt}, {Heffner-Wong}  \& {Tamanai}}{{Ferguson}
  et~al.}{2005}]{2005ApJ...623..585F}
{Ferguson} J.~W.,  {Alexander} D.~R.,  {Allard} F.,  {Barman} T.,  {Bodnarik}
  J.~G.,  {Hauschildt} P.~H.,  {Heffner-Wong} A.,   {Tamanai} A.,  2005,
  \mn@doi [\apj] {10.1086/428642}, \href
  {http://adsabs.harvard.edu/abs/2005ApJ...623..585F} {623, 585}

\bibitem[\protect\citeauthoryear{{Fontaine}, {Brassard}, {Charpinet}, {Green},
  {Randall}  \& {Van Grootel}}{{Fontaine} et~al.}{2012}]{2012A&A...539A..12F}
{Fontaine} G.,  {Brassard} P.,  {Charpinet} S.,  {Green} E.~M.,  {Randall}
  S.~K.,   {Van Grootel} V.,  2012, \mn@doi [\aap]
  {10.1051/0004-6361/201118220}, \href
  {http://adsabs.harvard.edu/abs/2012A%26A...539A..12F} {539, A12}

\bibitem[\protect\citeauthoryear{{Geier} \& {Heber}}{{Geier} \&
  {Heber}}{2012}]{2012A&A...543A.149G}
{Geier} S.,  {Heber} U.,  2012, \mn@doi [\aap] {10.1051/0004-6361/201219463},
  \href {http://adsabs.harvard.edu/abs/2012A%26A...543A.149G} {543, A149}

\bibitem[\protect\citeauthoryear{{Geier}, {Classen}  \& {Heber}}{{Geier}
  et~al.}{2011}]{2011ApJ...733L..13G}
{Geier} S.,  {Classen} L.,   {Heber} U.,  2011, \mn@doi [\apjl]
  {10.1088/2041-8205/733/1/L13}, \href
  {http://adsabs.harvard.edu/abs/2011ApJ...733L..13G} {733, L13}

\bibitem[\protect\citeauthoryear{{Geier}, {Heber}, {Heuser}, {Classen},
  {O'Toole}  \& {Edelmann}}{{Geier} et~al.}{2013a}]{2013A&A...551L...4G}
{Geier} S.,  {Heber} U.,  {Heuser} C.,  {Classen} L.,  {O'Toole} S.~J.,
  {Edelmann} H.,  2013a, \mn@doi [\aap] {10.1051/0004-6361/201220964}, \href
  {http://adsabs.harvard.edu/abs/2013A%26A...551L...4G} {551, L4}

\bibitem[\protect\citeauthoryear{{Geier}, {Heber}, {Edelmann}, {Morales-Rueda},
  {Kilkenny}, {O'Donoghue}, {Marsh}  \& {Copperwheat}}{{Geier}
  et~al.}{2013b}]{2013A&A...557A.122G}
{Geier} S.,  {Heber} U.,  {Edelmann} H.,  {Morales-Rueda} L.,  {Kilkenny} D.,
  {O'Donoghue} D.,  {Marsh} T.~R.,   {Copperwheat} C.,  2013b, \mn@doi [\aap]
  {10.1051/0004-6361/201322057}, \href
  {http://adsabs.harvard.edu/abs/2013A%26A...557A.122G} {557, A122}

\bibitem[\protect\citeauthoryear{{Gianninas}, {Kilic}, {Brown}, {Canton}  \&
  {Kenyon}}{{Gianninas} et~al.}{2015}]{2015ApJ...812..167G}
{Gianninas} A.,  {Kilic} M.,  {Brown} W.~R.,  {Canton} P.,   {Kenyon} S.~J.,
  2015, \mn@doi [\apj] {10.1088/0004-637X/812/2/167}, \href
  {http://adsabs.harvard.edu/abs/2015ApJ...812..167G} {812, 167}

\bibitem[\protect\citeauthoryear{{Giannone}}{{Giannone}}{1967}]{1967ZA.....65..226G}
{Giannone} P.,  1967, \zap, \href
  {http://adsabs.harvard.edu/abs/1967ZA.....65..226G} {65, 226}

\bibitem[\protect\citeauthoryear{{Giannone}, {Kohl}  \& {Weigert}}{{Giannone}
  et~al.}{1968}]{1968ZA.....68..107G}
{Giannone} P.,  {Kohl} K.,   {Weigert} A.,  1968, \zap, \href
  {http://adsabs.harvard.edu/abs/1968ZA.....68..107G} {68, 107}

\bibitem[\protect\citeauthoryear{{Gokhale}, {Peng}  \& {Frank}}{{Gokhale}
  et~al.}{2007}]{2007ApJ...655.1010G}
{Gokhale} V.,  {Peng} X.~M.,   {Frank} J.,  2007, \mn@doi [\apj]
  {10.1086/510119}, \href {http://adsabs.harvard.edu/abs/2007ApJ...655.1010G}
  {655, 1010}

\bibitem[\protect\citeauthoryear{{Graboske}, {Dewitt}, {Grossman}  \&
  {Cooper}}{{Graboske} et~al.}{1973}]{1973ApJ...181..457G}
{Graboske} H.~C.,  {Dewitt} H.~E.,  {Grossman} A.~S.,   {Cooper} M.~S.,  1973,
  \mn@doi [\apj] {10.1086/152062}, \href
  {http://adsabs.harvard.edu/abs/1973ApJ...181..457G} {181, 457}

\bibitem[\protect\citeauthoryear{{Grevesse} \& {Sauval}}{{Grevesse} \&
  {Sauval}}{1998}]{1998SSRv...85..161G}
{Grevesse} N.,  {Sauval} A.~J.,  1998, \mn@doi [\ssr]
  {10.1023/A:1005161325181}, \href
  {http://adsabs.harvard.edu/abs/1998SSRv...85..161G} {85, 161}

\bibitem[\protect\citeauthoryear{{Guerrero}, {Garc{\'{\i}}a-Berro}  \&
  {Isern}}{{Guerrero} et~al.}{2004}]{2004A&A...413..257G}
{Guerrero} J.,  {Garc{\'{\i}}a-Berro} E.,   {Isern} J.,  2004, \mn@doi [\aap]
  {10.1051/0004-6361:20031504}, \href
  {http://adsabs.harvard.edu/abs/2004A%26A...413..257G} {413, 257}

\bibitem[\protect\citeauthoryear{{Hall}, {Tout}, {Izzard}  \& {Keller}}{{Hall}
  et~al.}{2013}]{2013MNRAS.435.2048H}
{Hall} P.~D.,  {Tout} C.~A.,  {Izzard} R.~G.,   {Keller} D.,  2013, \mn@doi
  [\mnras] {10.1093/mnras/stt1422}, \href
  {http://adsabs.harvard.edu/abs/2013MNRAS.435.2048H} {435, 2048}

\bibitem[\protect\citeauthoryear{{Han}}{{Han}}{1998}]{1998MNRAS.296.1019H}
{Han} Z.,  1998, \mn@doi [\mnras] {10.1046/j.1365-8711.1998.01475.x}, \href
  {http://adsabs.harvard.edu/abs/1998MNRAS.296.1019H} {296, 1019}

\bibitem[\protect\citeauthoryear{{Han}, {Podsiadlowski}  \& {Eggleton}}{{Han}
  et~al.}{1994}]{1994MNRAS.270..121H}
{Han} Z.,  {Podsiadlowski} P.,   {Eggleton} P.~P.,  1994, \mnras, \href
  {http://adsabs.harvard.edu/abs/1994MNRAS.270..121H} {270, 121}

\bibitem[\protect\citeauthoryear{{Han}, {Podsiadlowski}, {Maxted}, {Marsh}  \&
  {Ivanova}}{{Han} et~al.}{2002}]{2002MNRAS.336..449H}
{Han} Z.,  {Podsiadlowski} P.,  {Maxted} P.~F.~L.,  {Marsh} T.~R.,   {Ivanova}
  N.,  2002, \mn@doi [\mnras] {10.1046/j.1365-8711.2002.05752.x}, \href
  {http://adsabs.harvard.edu/abs/2002MNRAS.336..449H} {336, 449}

\bibitem[\protect\citeauthoryear{{Han}, {Podsiadlowski}, {Maxted}  \&
  {Marsh}}{{Han} et~al.}{2003}]{2003MNRAS.341..669H}
{Han} Z.,  {Podsiadlowski} P.,  {Maxted} P.~F.~L.,   {Marsh} T.~R.,  2003,
  \mn@doi [\mnras] {10.1046/j.1365-8711.2003.06451.x}, \href
  {http://adsabs.harvard.edu/abs/2003MNRAS.341..669H} {341, 669}

\bibitem[\protect\citeauthoryear{{Heber}}{{Heber}}{1986}]{1986A&A...155...33H}
{Heber} U.,  1986, \aap, \href
  {http://adsabs.harvard.edu/abs/1986A%26A...155...33H} {155, 33}

\bibitem[\protect\citeauthoryear{{Heber}}{{Heber}}{2009}]{2009ARA&A..47..211H}
{Heber} U.,  2009, \mn@doi [\araa] {10.1146/annurev-astro-082708-101836}, \href
  {http://adsabs.harvard.edu/abs/2009ARA%26A..47..211H} {47, 211}

\bibitem[\protect\citeauthoryear{{Heber}}{{Heber}}{2016}]{2016PASP..128h2001H}
{Heber} U.,  2016, \mn@doi [\pasp] {10.1088/1538-3873/128/966/082001}, \href
  {http://adsabs.harvard.edu/abs/2016PASP..128h2001H} {128, 082001}

\bibitem[\protect\citeauthoryear{{Heber}, {Edelmann}, {Lisker}  \&
  {Napiwotzki}}{{Heber} et~al.}{2003}]{2003A&A...411L.477H}
{Heber} U.,  {Edelmann} H.,  {Lisker} T.,   {Napiwotzki} R.,  2003, \mn@doi
  [\aap] {10.1051/0004-6361:20031553}, \href
  {http://adsabs.harvard.edu/abs/2003A%26A...411L.477H} {411, L477}

\bibitem[\protect\citeauthoryear{{Hermes}, {Montgomery}, {Winget}, {Brown},
  {Kilic}  \& {Kenyon}}{{Hermes} et~al.}{2012}]{2012ApJ...750L..28H}
{Hermes} J.~J.,  {Montgomery} M.~H.,  {Winget} D.~E.,  {Brown} W.~R.,  {Kilic}
  M.,   {Kenyon} S.~J.,  2012, \mn@doi [\apjl] {10.1088/2041-8205/750/2/L28},
  \href {http://adsabs.harvard.edu/abs/2012ApJ...750L..28H} {750, L28}

\bibitem[\protect\citeauthoryear{{Herwig}}{{Herwig}}{2000}]{2000A&A...360..952H}
{Herwig} F.,  2000, \aap, \href
  {http://adsabs.harvard.edu/abs/2000A%26A...360..952H} {360, 952}

\bibitem[\protect\citeauthoryear{{Hix}, {Khokhlov}, {Wheeler}  \&
  {Thielemann}}{{Hix} et~al.}{1998}]{1998ApJ...503..332H}
{Hix} W.~R.,  {Khokhlov} A.~M.,  {Wheeler} J.~C.,   {Thielemann} F.-K.,  1998,
  \mn@doi [\apj] {10.1086/305968}, \href
  {http://adsabs.harvard.edu/abs/1998ApJ...503..332H} {503, 332}

\bibitem[\protect\citeauthoryear{{Hu}, {Glebbeek}, {Thoul}, {Dupret},
  {Stancliffe}, {Nelemans}  \& {Aerts}}{{Hu}
  et~al.}{2010}]{2010A&A...511A..87H}
{Hu} H.,  {Glebbeek} E.,  {Thoul} A.~A.,  {Dupret} M.-A.,  {Stancliffe} R.~J.,
  {Nelemans} G.,   {Aerts} C.,  2010, \mn@doi [\aap]
  {10.1051/0004-6361/200912290}, \href
  {http://adsabs.harvard.edu/abs/2010A%26A...511A..87H} {511, A87}

\bibitem[\protect\citeauthoryear{{Iben}}{{Iben}}{1988}]{1988ApJ...324..355I}
{Iben} Jr. I.,  1988, \mn@doi [\apj] {10.1086/165900}, \href
  {http://adsabs.harvard.edu/abs/1988ApJ...324..355I} {324, 355}

\bibitem[\protect\citeauthoryear{{Iben}}{{Iben}}{1990}]{1990ApJ...353..215I}
{Iben} Jr. I.,  1990, \mn@doi [\apj] {10.1086/168609}, \href
  {http://adsabs.harvard.edu/abs/1990ApJ...353..215I} {353, 215}

\bibitem[\protect\citeauthoryear{{Iben} \& {Tutukov}}{{Iben} \&
  {Tutukov}}{1984}]{1984ApJS...54..335I}
{Iben} Jr. I.,  {Tutukov} A.~V.,  1984, \mn@doi [\apjs] {10.1086/190932}, \href
  {http://adsabs.harvard.edu/abs/1984ApJS...54..335I} {54, 335}

\bibitem[\protect\citeauthoryear{{Iben} \& {Tutukov}}{{Iben} \&
  {Tutukov}}{1986a}]{1986ApJ...311..742I}
{Iben} Jr. I.,  {Tutukov} A.~V.,  1986a, \mn@doi [\apj] {10.1086/164812}, \href
  {http://adsabs.harvard.edu/abs/1986ApJ...311..742I} {311, 742}

\bibitem[\protect\citeauthoryear{{Iben} \& {Tutukov}}{{Iben} \&
  {Tutukov}}{1986b}]{1986ApJ...311..753I}
{Iben} Jr. I.,  {Tutukov} A.~V.,  1986b, \mn@doi [\apj] {10.1086/164813}, \href
  {http://adsabs.harvard.edu/abs/1986ApJ...311..753I} {311, 753}

\bibitem[\protect\citeauthoryear{{Iben} \& {Webbink}}{{Iben} \&
  {Webbink}}{1987}]{1987fbs..conf..401I}
{Iben} Jr. I.,  {Webbink} R.~F.,  1987, in {Philip} A.~G.~D.,  {Hayes} D.~S.,
  {Liebert} J.~W.,  eds, IAU Colloq.\ 95, Second Conference on Faint Blue
  Stars. {Davis Press}, {Schenectady}, p.~401

\bibitem[\protect\citeauthoryear{{Iglesias} \& {Rogers}}{{Iglesias} \&
  {Rogers}}{1996}]{1996ApJ...464..943I}
{Iglesias} C.~A.,  {Rogers} F.~J.,  1996, \mn@doi [\apj] {10.1086/177381},
  \href {http://adsabs.harvard.edu/abs/1996ApJ...464..943I} {464, 943}

\bibitem[\protect\citeauthoryear{{Itoh}, {Totsuji}, {Ichimaru}  \&
  {Dewitt}}{{Itoh} et~al.}{1979}]{1979ApJ...234.1079I}
{Itoh} N.,  {Totsuji} H.,  {Ichimaru} S.,   {Dewitt} H.~E.,  1979, \mn@doi
  [\apj] {10.1086/157590}, \href
  {http://adsabs.harvard.edu/abs/1979ApJ...234.1079I} {234, 1079}

\bibitem[\protect\citeauthoryear{{Itoh}, {Hayashi}, {Nishikawa}  \&
  {Kohyama}}{{Itoh} et~al.}{1996}]{1996ApJS..102..411I}
{Itoh} N.,  {Hayashi} H.,  {Nishikawa} A.,   {Kohyama} Y.,  1996, \mn@doi
  [\apjs] {10.1086/192264}, \href
  {http://adsabs.harvard.edu/abs/1996ApJS..102..411I} {102, 411}

\bibitem[\protect\citeauthoryear{{Ivanova}}{{Ivanova}}{2011}]{2011ApJ...730...76I}
{Ivanova} N.,  2011, \mn@doi [\apj] {10.1088/0004-637X/730/2/76}, \href
  {http://adsabs.harvard.edu/abs/2011ApJ...730...76I} {730, 76}

\bibitem[\protect\citeauthoryear{{Justham}, {Podsiadlowski}  \&
  {Han}}{{Justham} et~al.}{2011}]{2011MNRAS.410..984J}
{Justham} S.,  {Podsiadlowski} P.,   {Han} Z.,  2011, \mn@doi [\mnras]
  {10.1111/j.1365-2966.2010.17497.x}, \href
  {http://adsabs.harvard.edu/abs/2011MNRAS.410..984J} {410, 984}

\bibitem[\protect\citeauthoryear{{Kaplan}, {Bildsten}  \& {Steinfadt}}{{Kaplan}
  et~al.}{2012}]{2012ApJ...758...64K}
{Kaplan} D.~L.,  {Bildsten} L.,   {Steinfadt} J.~D.~R.,  2012, \mn@doi [\apj]
  {10.1088/0004-637X/758/1/64}, \href
  {http://adsabs.harvard.edu/abs/2012ApJ...758...64K} {758, 64}

\bibitem[\protect\citeauthoryear{{Kilic}, {Brown}, {Allende Prieto},
  {Ag{\"u}eros}, {Heinke}  \& {Kenyon}}{{Kilic}
  et~al.}{2011}]{2011ApJ...727....3K}
{Kilic} M.,  {Brown} W.~R.,  {Allende Prieto} C.,  {Ag{\"u}eros} M.~A.,
  {Heinke} C.,   {Kenyon} S.~J.,  2011, \mn@doi [\apj]
  {10.1088/0004-637X/727/1/3}, \href
  {http://adsabs.harvard.edu/abs/2011ApJ...727....3K} {727, 3}

\bibitem[\protect\citeauthoryear{{Kilic}, {Brown}, {Allende Prieto}, {Kenyon},
  {Heinke}, {Ag{\"u}eros}  \& {Kleinman}}{{Kilic}
  et~al.}{2012}]{2012ApJ...751..141K}
{Kilic} M.,  {Brown} W.~R.,  {Allende Prieto} C.,  {Kenyon} S.~J.,  {Heinke}
  C.~O.,  {Ag{\"u}eros} M.~A.,   {Kleinman} S.~J.,  2012, \mn@doi [\apj]
  {10.1088/0004-637X/751/2/141}, \href
  {http://adsabs.harvard.edu/abs/2012ApJ...751..141K} {751, 141}

\bibitem[\protect\citeauthoryear{{Kremer}, {Sepinsky}  \& {Kalogera}}{{Kremer}
  et~al.}{2015}]{2015ApJ...806...76K}
{Kremer} K.,  {Sepinsky} J.,   {Kalogera} V.,  2015, \mn@doi [\apj]
  {10.1088/0004-637X/806/1/76}, \href
  {http://adsabs.harvard.edu/abs/2015ApJ...806...76K} {806, 76}

\bibitem[\protect\citeauthoryear{{Lambert} \& {Rao}}{{Lambert} \&
  {Rao}}{1994}]{1994JApA...15...47L}
{Lambert} D.~L.,  {Rao} N.~K.,  1994, \mn@doi [Journal of Astrophysics and
  Astronomy] {10.1007/BF03010404}, \href
  {http://adsabs.harvard.edu/abs/1994JApA...15...47L} {15, 47}

\bibitem[\protect\citeauthoryear{{Lisker}, {Heber}, {Napiwotzki}, {Christlieb},
  {Han}, {Homeier}  \& {Reimers}}{{Lisker} et~al.}{2005}]{2005A&A...430..223L}
{Lisker} T.,  {Heber} U.,  {Napiwotzki} R.,  {Christlieb} N.,  {Han} Z.,
  {Homeier} D.,   {Reimers} D.,  2005, \mn@doi [\aap]
  {10.1051/0004-6361:20040232}, \href
  {http://adsabs.harvard.edu/abs/2005A%26A...430..223L} {430, 223}

\bibitem[\protect\citeauthoryear{{Lor{\'e}n-Aguilar}, {Isern}  \&
  {Garc{\'{\i}}a-Berro}}{{Lor{\'e}n-Aguilar}
  et~al.}{2009}]{2009A&A...500.1193L}
{Lor{\'e}n-Aguilar} P.,  {Isern} J.,   {Garc{\'{\i}}a-Berro} E.,  2009, \mn@doi
  [\aap] {10.1051/0004-6361/200811060}, \href
  {http://adsabs.harvard.edu/abs/2009A%26A...500.1193L} {500, 1193}

\bibitem[\protect\citeauthoryear{{Marsh}, {Dhillon}  \& {Duck}}{{Marsh}
  et~al.}{1995}]{1995MNRAS.275..828M}
{Marsh} T.~R.,  {Dhillon} V.~S.,   {Duck} S.~R.,  1995, \mn@doi [\mnras]
  {10.1093/mnras/275.3.828}, \href
  {http://adsabs.harvard.edu/abs/1995MNRAS.275..828M} {275, 828}

\bibitem[\protect\citeauthoryear{{Marsh}, {Nelemans}  \& {Steeghs}}{{Marsh}
  et~al.}{2004}]{2004MNRAS.350..113M}
{Marsh} T.~R.,  {Nelemans} G.,   {Steeghs} D.,  2004, \mn@doi [\mnras]
  {10.1111/j.1365-2966.2004.07564.x}, \href
  {http://adsabs.harvard.edu/abs/2004MNRAS.350..113M} {350, 113}

\bibitem[\protect\citeauthoryear{{Maxted}, {Heber}, {Marsh}  \&
  {North}}{{Maxted} et~al.}{2001}]{2001MNRAS.326.1391M}
{Maxted} P.~F.~L.,  {Heber} U.,  {Marsh} T.~R.,   {North} R.~C.,  2001, \mn@doi
  [\mnras] {10.1111/j.1365-2966.2001.04714.x}, \href
  {http://adsabs.harvard.edu/abs/2001MNRAS.326.1391M} {326, 1391}

\bibitem[\protect\citeauthoryear{{Maxted} et~al.,}{{Maxted}
  et~al.}{2013}]{2013Natur.498..463M}
{Maxted} P.~F.~L.  et~al., 2013, \mn@doi [\nat] {10.1038/nature12192}, \href
  {http://adsabs.harvard.edu/abs/2013Natur.498..463M} {498, 463}

\bibitem[\protect\citeauthoryear{{Mengel}, {Norris}  \& {Gross}}{{Mengel}
  et~al.}{1976}]{1976ApJ...204..488M}
{Mengel} J.~G.,  {Norris} J.,   {Gross} P.~G.,  1976, \mn@doi [\apj]
  {10.1086/154193}, \href {http://adsabs.harvard.edu/abs/1976ApJ...204..488M}
  {204, 488}

\bibitem[\protect\citeauthoryear{{Motl}, {Frank}, {Tohline}  \&
  {D'Souza}}{{Motl} et~al.}{2007}]{2007ApJ...670.1314M}
{Motl} P.~M.,  {Frank} J.,  {Tohline} J.~E.,   {D'Souza} M.~C.~R.,  2007,
  \mn@doi [\apj] {10.1086/522076}, \href
  {http://adsabs.harvard.edu/abs/2007ApJ...670.1314M} {670, 1314}

\bibitem[\protect\citeauthoryear{{Nelemans} \& {Tout}}{{Nelemans} \&
  {Tout}}{2005}]{2005MNRAS.356..753N}
{Nelemans} G.,  {Tout} C.~A.,  2005, \mn@doi [\mnras]
  {10.1111/j.1365-2966.2004.08496.x}, \href
  {http://adsabs.harvard.edu/abs/2005MNRAS.356..753N} {356, 753}

\bibitem[\protect\citeauthoryear{{Nelemans}, {Verbunt}, {Yungelson}  \&
  {Portegies Zwart}}{{Nelemans} et~al.}{2000}]{2000A&A...360.1011N}
{Nelemans} G.,  {Verbunt} F.,  {Yungelson} L.~R.,   {Portegies Zwart} S.~F.,
  2000, \aap, \href {http://adsabs.harvard.edu/abs/2000A%26A...360.1011N} {360,
  1011}

\bibitem[\protect\citeauthoryear{{Nelemans}, {Yungelson}, {Portegies Zwart}  \&
  {Verbunt}}{{Nelemans} et~al.}{2001a}]{2001A&A...365..491N}
{Nelemans} G.,  {Yungelson} L.~R.,  {Portegies Zwart} S.~F.,   {Verbunt} F.,
  2001a, \mn@doi [\aap] {10.1051/0004-6361:20000147}, \href
  {http://adsabs.harvard.edu/abs/2001A%26A...365..491N} {365, 491}

\bibitem[\protect\citeauthoryear{{Nelemans}, {Portegies Zwart}, {Verbunt}  \&
  {Yungelson}}{{Nelemans} et~al.}{2001b}]{2001A&A...368..939N}
{Nelemans} G.,  {Portegies Zwart} S.~F.,  {Verbunt} F.,   {Yungelson} L.~R.,
  2001b, \mn@doi [\aap] {10.1051/0004-6361:20010049}, \href
  {http://adsabs.harvard.edu/abs/2001A%26A...368..939N} {368, 939}

\bibitem[\protect\citeauthoryear{{Nelemans} et~al.,}{{Nelemans}
  et~al.}{2005}]{2005A&A...440.1087N}
{Nelemans} G.  et~al., 2005, \mn@doi [\aap] {10.1051/0004-6361:20053174}, \href
  {http://adsabs.harvard.edu/abs/2005A%26A...440.1087N} {440, 1087}

\bibitem[\protect\citeauthoryear{{N{\'e}meth}, {Kawka}  \&
  {Vennes}}{{N{\'e}meth} et~al.}{2012}]{2012MNRAS.427.2180N}
{N{\'e}meth} P.,  {Kawka} A.,   {Vennes} S.,  2012, \mn@doi [\mnras]
  {10.1111/j.1365-2966.2012.22009.x}, \href
  {http://adsabs.harvard.edu/abs/2012MNRAS.427.2180N} {427, 2180}

\bibitem[\protect\citeauthoryear{{Nomoto} \& {Iben}}{{Nomoto} \&
  {Iben}}{1985}]{1985ApJ...297..531N}
{Nomoto} K.,  {Iben} Jr. I.,  1985, \mn@doi [\apj] {10.1086/163547}, \href
  {http://adsabs.harvard.edu/abs/1985ApJ...297..531N} {297, 531}

\bibitem[\protect\citeauthoryear{{Paczy{\'n}ski}}{{Paczy{\'n}ski}}{1967}]{1967AcA....17..287P}
{Paczy{\'n}ski} B.,  1967, \actaa, \href
  {http://adsabs.harvard.edu/abs/1967AcA....17..287P} {17, 287}

\bibitem[\protect\citeauthoryear{{Pakmor}, {Kromer}, {Taubenberger}  \&
  {Springel}}{{Pakmor} et~al.}{2013}]{2013ApJ...770L...8P}
{Pakmor} R.,  {Kromer} M.,  {Taubenberger} S.,   {Springel} V.,  2013, \mn@doi
  [\apjl] {10.1088/2041-8205/770/1/L8}, \href
  {http://adsabs.harvard.edu/abs/2013ApJ...770L...8P} {770, L8}

\bibitem[\protect\citeauthoryear{{Parsons}, {Marsh}, {G{\"a}nsicke}, {Drake}
  \& {Koester}}{{Parsons} et~al.}{2011}]{2011ApJ...735L..30P}
{Parsons} S.~G.,  {Marsh} T.~R.,  {G{\"a}nsicke} B.~T.,  {Drake} A.~J.,
  {Koester} D.,  2011, \mn@doi [\apjl] {10.1088/2041-8205/735/2/L30}, \href
  {http://adsabs.harvard.edu/abs/2011ApJ...735L..30P} {735, L30}

\bibitem[\protect\citeauthoryear{{Paxton}, {Bildsten}, {Dotter}, {Herwig},
  {Lesaffre}  \& {Timmes}}{{Paxton} et~al.}{2011}]{2011ApJS..192....3P}
{Paxton} B.,  {Bildsten} L.,  {Dotter} A.,  {Herwig} F.,  {Lesaffre} P.,
  {Timmes} F.,  2011, \mn@doi [\apjs] {10.1088/0067-0049/192/1/3}, \href
  {http://adsabs.harvard.edu/abs/2011ApJS..192....3P} {192, 3}

\bibitem[\protect\citeauthoryear{{Paxton} et~al.,}{{Paxton}
  et~al.}{2013}]{2013ApJS..208....4P}
{Paxton} B.  et~al., 2013, \mn@doi [\apjs] {10.1088/0067-0049/208/1/4}, \href
  {http://adsabs.harvard.edu/abs/2013ApJS..208....4P} {208, 4}

\bibitem[\protect\citeauthoryear{{Paxton} et~al.,}{{Paxton}
  et~al.}{2015}]{2015ApJS..220...15P}
{Paxton} B.  et~al., 2015, \mn@doi [\apjs] {10.1088/0067-0049/220/1/15}, \href
  {http://adsabs.harvard.edu/abs/2015ApJS..220...15P} {220, 15}

\bibitem[\protect\citeauthoryear{{Politano}, {Taam}, {van der Sluys}  \&
  {Willems}}{{Politano} et~al.}{2008}]{2008ApJ...687L..99P}
{Politano} M.,  {Taam} R.~E.,  {van der Sluys} M.,   {Willems} B.,  2008,
  \mn@doi [\apjl] {10.1086/593328}, \href
  {http://adsabs.harvard.edu/abs/2008ApJ...687L..99P} {687, L99}

\bibitem[\protect\citeauthoryear{{Potekhin} \& {Chabrier}}{{Potekhin} \&
  {Chabrier}}{2010}]{2010CoPP...50...82P}
{Potekhin} A.~Y.,  {Chabrier} G.,  2010, \mn@doi [Contributions to Plasma
  Physics] {10.1002/ctpp.201010017}, \href
  {http://adsabs.harvard.edu/abs/2010CoPP...50...82P} {50, 82}

\bibitem[\protect\citeauthoryear{{Raskin}, {Scannapieco}, {Fryer},
  {Rockefeller}  \& {Timmes}}{{Raskin} et~al.}{2012}]{2012ApJ...746...62R}
{Raskin} C.,  {Scannapieco} E.,  {Fryer} C.,  {Rockefeller} G.,   {Timmes}
  F.~X.,  2012, \mn@doi [\apj] {10.1088/0004-637X/746/1/62}, \href
  {http://adsabs.harvard.edu/abs/2012ApJ...746...62R} {746, 62}

\bibitem[\protect\citeauthoryear{{Rogers} \& {Nayfonov}}{{Rogers} \&
  {Nayfonov}}{2002}]{2002ApJ...576.1064R}
{Rogers} F.~J.,  {Nayfonov} A.,  2002, \mn@doi [\apj] {10.1086/341894}, \href
  {http://adsabs.harvard.edu/abs/2002ApJ...576.1064R} {576, 1064}

\bibitem[\protect\citeauthoryear{{Saffer}, {Liebert}  \& {Olszewski}}{{Saffer}
  et~al.}{1988}]{1988ApJ...334..947S}
{Saffer} R.~A.,  {Liebert} J.,   {Olszewski} E.~W.,  1988, \mn@doi [\apj]
  {10.1086/166888}, \href {http://adsabs.harvard.edu/abs/1988ApJ...334..947S}
  {334, 947}

\bibitem[\protect\citeauthoryear{{Saio} \& {Jeffery}}{{Saio} \&
  {Jeffery}}{2000}]{2000MNRAS.313..671S}
{Saio} H.,  {Jeffery} C.~S.,  2000, \mn@doi [\mnras]
  {10.1046/j.1365-8711.2000.03221.x}, \href
  {http://adsabs.harvard.edu/abs/2000MNRAS.313..671S} {313, 671}

\bibitem[\protect\citeauthoryear{{Saio} \& {Nomoto}}{{Saio} \&
  {Nomoto}}{1998}]{1998ApJ...500..388S}
{Saio} H.,  {Nomoto} K.,  1998, \mn@doi [\apj] {10.1086/305696}, \href
  {http://adsabs.harvard.edu/abs/1998ApJ...500..388S} {500, 388}

\bibitem[\protect\citeauthoryear{{Sarna}, {Marks}  \& {Connon Smith}}{{Sarna}
  et~al.}{1996}]{1996MNRAS.279...88S}
{Sarna} M.~J.,  {Marks} P.~B.,   {Connon Smith} R.,  1996, \mn@doi [\mnras]
  {10.1093/mnras/279.1.88}, \href
  {http://adsabs.harvard.edu/abs/1996MNRAS.279...88S} {279, 88}

\bibitem[\protect\citeauthoryear{{Sato}, {Nakasato}, {Tanikawa}, {Nomoto},
  {Maeda}  \& {Hachisu}}{{Sato} et~al.}{2015}]{2015ApJ...807..105S}
{Sato} Y.,  {Nakasato} N.,  {Tanikawa} A.,  {Nomoto} K.,  {Maeda} K.,
  {Hachisu} I.,  2015, \mn@doi [\apj] {10.1088/0004-637X/807/1/105}, \href
  {http://adsabs.harvard.edu/abs/2015ApJ...807..105S} {807, 105}

\bibitem[\protect\citeauthoryear{{Saumon}, {Chabrier}  \& {van Horn}}{{Saumon}
  et~al.}{1995}]{1995ApJS...99..713S}
{Saumon} D.,  {Chabrier} G.,   {van Horn} H.~M.,  1995, \mn@doi [\apjs]
  {10.1086/192204}, \href {http://adsabs.harvard.edu/abs/1995ApJS...99..713S}
  {99, 713}

\bibitem[\protect\citeauthoryear{{Schwab}, {Shen}, {Quataert}, {Dan}  \&
  {Rosswog}}{{Schwab} et~al.}{2012}]{2012MNRAS.427..190S}
{Schwab} J.,  {Shen} K.~J.,  {Quataert} E.,  {Dan} M.,   {Rosswog} S.,  2012,
  \mn@doi [\mnras] {10.1111/j.1365-2966.2012.21993.x}, \href
  {http://adsabs.harvard.edu/abs/2012MNRAS.427..190S} {427, 190}

\bibitem[\protect\citeauthoryear{{Sepinsky} \& {Kalogera}}{{Sepinsky} \&
  {Kalogera}}{2014}]{2014ApJ...785..157S}
{Sepinsky} J.~F.,  {Kalogera} V.,  2014, \mn@doi [\apj]
  {10.1088/0004-637X/785/2/157}, \href
  {http://adsabs.harvard.edu/abs/2014ApJ...785..157S} {785, 157}

\bibitem[\protect\citeauthoryear{{Shen}}{{Shen}}{2015}]{2015ApJ...805L...6S}
{Shen} K.~J.,  2015, \mn@doi [\apjl] {10.1088/2041-8205/805/1/L6}, \href
  {http://adsabs.harvard.edu/abs/2015ApJ...805L...6S} {805, L6}

\bibitem[\protect\citeauthoryear{{Shen}, {Bildsten}, {Kasen}  \&
  {Quataert}}{{Shen} et~al.}{2012}]{2012ApJ...748...35S}
{Shen} K.~J.,  {Bildsten} L.,  {Kasen} D.,   {Quataert} E.,  2012, \mn@doi
  [\apj] {10.1088/0004-637X/748/1/35}, \href
  {http://adsabs.harvard.edu/abs/2012ApJ...748...35S} {748, 35}

\bibitem[\protect\citeauthoryear{{Shen}, {Guillochon}  \& {Foley}}{{Shen}
  et~al.}{2013}]{2013ApJ...770L..35S}
{Shen} K.~J.,  {Guillochon} J.,   {Foley} R.~J.,  2013, \mn@doi [\apjl]
  {10.1088/2041-8205/770/2/L35}, \href
  {http://adsabs.harvard.edu/abs/2013ApJ...770L..35S} {770, L35}

\bibitem[\protect\citeauthoryear{{Staff} et~al.,}{{Staff}
  et~al.}{2012}]{2012ApJ...757...76S}
{Staff} J.~E.  et~al., 2012, \mn@doi [\apj] {10.1088/0004-637X/757/1/76}, \href
  {http://adsabs.harvard.edu/abs/2012ApJ...757...76S} {757, 76}

\bibitem[\protect\citeauthoryear{{Stroeer}, {Heber}, {Lisker}, {Napiwotzki},
  {Dreizler}, {Christlieb}  \& {Reimers}}{{Stroeer}
  et~al.}{2007}]{2007A&A...462..269S}
{Stroeer} A.,  {Heber} U.,  {Lisker} T.,  {Napiwotzki} R.,  {Dreizler} S.,
  {Christlieb} N.,   {Reimers} D.,  2007, \mn@doi [\aap]
  {10.1051/0004-6361:20065564}, \href
  {http://adsabs.harvard.edu/abs/2007A%26A...462..269S} {462, 269}

\bibitem[\protect\citeauthoryear{{Sweigart}}{{Sweigart}}{1997}]{1997ApJ...474L..23S}
{Sweigart} A.~V.,  1997, \mn@doi [\apjl] {10.1086/310414}, \href
  {http://adsabs.harvard.edu/abs/1997ApJ...474L..23S} {474, L23}

\bibitem[\protect\citeauthoryear{{Timmes} \& {Swesty}}{{Timmes} \&
  {Swesty}}{2000}]{2000ApJS..126..501T}
{Timmes} F.~X.,  {Swesty} F.~D.,  2000, \mn@doi [\apjs] {10.1086/313304}, \href
  {http://adsabs.harvard.edu/abs/2000ApJS..126..501T} {126, 501}

\bibitem[\protect\citeauthoryear{{Tutukov} \& {Yungelson}}{{Tutukov} \&
  {Yungelson}}{1979}]{1979AcA....29..665T}
{Tutukov} A.~V.,  {Yungelson} L.~R.,  1979, \actaa, \href
  {http://adsabs.harvard.edu/abs/1979AcA....29..665T} {29, 665}

\bibitem[\protect\citeauthoryear{{Vos}, {{\O}stensen}, {N{\'e}meth}, {Green},
  {Heber}  \& {Van Winckel}}{{Vos} et~al.}{2013}]{2013A&A...559A..54V}
{Vos} J.,  {{\O}stensen} R.~H.,  {N{\'e}meth} P.,  {Green} E.~M.,  {Heber} U.,
   {Van Winckel} H.,  2013, \mn@doi [\aap] {10.1051/0004-6361/201322200}, \href
  {http://adsabs.harvard.edu/abs/2013A%26A...559A..54V} {559, A54}

\bibitem[\protect\citeauthoryear{{Webbink}}{{Webbink}}{1984}]{1984ApJ...277..355W}
{Webbink} R.~F.,  1984, \mn@doi [\apj] {10.1086/161701}, \href
  {http://adsabs.harvard.edu/abs/1984ApJ...277..355W} {277, 355}

\bibitem[\protect\citeauthoryear{{Woods}, {Ivanova}, {van der Sluys}  \&
  {Chaichenets}}{{Woods} et~al.}{2012}]{2012ApJ...744...12W}
{Woods} T.~E.,  {Ivanova} N.,  {van der Sluys} M.~V.,   {Chaichenets} S.,
  2012, \mn@doi [\apj] {10.1088/0004-637X/744/1/12}, \href
  {http://adsabs.harvard.edu/abs/2012ApJ...744...12W} {744, 12}

\bibitem[\protect\citeauthoryear{{Yu} \& {Jeffery}}{{Yu} \&
  {Jeffery}}{2011}]{2011MNRAS.417.1392Y}
{Yu} S.,  {Jeffery} C.~S.,  2011, \mn@doi [\mnras]
  {10.1111/j.1365-2966.2011.19352.x}, \href
  {http://adsabs.harvard.edu/abs/2011MNRAS.417.1392Y} {417, 1392}

\bibitem[\protect\citeauthoryear{{Yu} \& {Li}}{{Yu} \&
  {Li}}{2009}]{2009A&A...503..151Y}
{Yu} S.,  {Li} L.,  2009, \mn@doi [\aap] {10.1051/0004-6361/200809454}, \href
  {http://adsabs.harvard.edu/abs/2009A%26A...503..151Y} {503, 151}

\bibitem[\protect\citeauthoryear{{Zhang} \& {Jeffery}}{{Zhang} \&
  {Jeffery}}{2012a}]{2012MNRAS.419..452Z}
{Zhang} X.,  {Jeffery} C.~S.,  2012a, \mn@doi [\mnras]
  {10.1111/j.1365-2966.2011.19711.x}, \href
  {http://adsabs.harvard.edu/abs/2012MNRAS.419..452Z} {419, 452}

\bibitem[\protect\citeauthoryear{{Zhang} \& {Jeffery}}{{Zhang} \&
  {Jeffery}}{2012b}]{2012MNRAS.426L..81Z}
{Zhang} X.,  {Jeffery} C.~S.,  2012b, \mn@doi [\mnras]
  {10.1111/j.1745-3933.2012.01330.x}, \href
  {http://adsabs.harvard.edu/abs/2012MNRAS.426L..81Z} {426, L81}

\bibitem[\protect\citeauthoryear{{Zhu}, {Pakmor}, {van Kerkwijk}  \&
  {Chang}}{{Zhu} et~al.}{2015}]{2015ApJ...806L...1Z}
{Zhu} C.,  {Pakmor} R.,  {van Kerkwijk} M.~H.,   {Chang} P.,  2015, \mn@doi
  [\apjl] {10.1088/2041-8205/806/1/L1}, \href
  {http://adsabs.harvard.edu/abs/2015ApJ...806L...1Z} {806, L1}

\bibitem[\protect\citeauthoryear{{van Winckel}}{{van
  Winckel}}{2003}]{2003ARA&A..41..391V}
{van Winckel} H.,  2003, \mn@doi [\araa]
  {10.1146/annurev.astro.41.071601.170018}, \href
  {http://adsabs.harvard.edu/abs/2003ARA%26A..41..391V} {41, 391}

\bibitem[\protect\citeauthoryear{{van der Sluys}, {Verbunt}  \& {Pols}}{{van
  der Sluys} et~al.}{2006}]{2006A&A...460..209V}
{van der Sluys} M.~V.,  {Verbunt} F.,   {Pols} O.~R.,  2006, \mn@doi [\aap]
  {10.1051/0004-6361:20065066}, \href
  {http://adsabs.harvard.edu/abs/2006A%26A...460..209V} {460, 209}

\makeatother
\end{thebibliography}

\appendix
\section{\textsc{mesa/star} inlist}\label{sec:mesa_inlist}
The \textsc{mesa/star} inlist parameters that differ from the defaults are as follows:
{\small
\begin{verbatim}
&star_job
   change_net = .true.                                                         
   new_net_name = 'o18_and_ne22.net'

   kappa_file_prefix = 'gs98'                                        
   kappa_CO_prefix = 'gs98_co' 
/

&controls
   use_Type2_opacities = .true. 
   initial_z = 0.02
   Zbase = 0.02

   overshoot_f_<above/below>_<region> = 0.014
   overshoot_f0_<above/below>_<region> = 0.004

   which_atm_option = 'Eddington_grey'

   he_core_boundary_h1_fraction = 0.1
/
\end{verbatim}
}

\section{Observed hot subdwarfs}
\begin{table*}
\centering
\begin{minipage}{10cm}
\caption{Atmospheric parameters of $48$ apparently isolated hot subdwarfs.}
\label{tab:sample}
\begin{tabular}{lrrrrl}
\toprule
  System & \multicolumn{1}{c}{$\dfrac{T_{\rmn{eff}}}{\rmn{kK}}$} & \multicolumn{1}{c}{$\log_{10}\left(\dfrac{g}{\rmn{cm}\,\rmn{s}^{-2}}\right)$} & \multicolumn{1}{c}{$\log_{10}\left(\dfrac{n_{\rmn{He}}}{n_{\rmn{H}}}\right)$} & \multicolumn{1}{c}{$\dfrac{v_{\rmn{rot}} \sin i}{\rmn{km\,s^{-1}}}$} & Source \\
\midrule
EC\,$20106{-}5248$ & $24.5$ & $5.25$ & $-2.77$ & $7.8 \pm 1.0$ & {\citet{2013A&A...557A.122G}} \\ 
BD${+}48^{\circ}\,2721$ & $24.8$ & $5.38$ & $-2.23$ & $4.7 \pm  1.4$ & {\citet{2013A&A...557A.122G}} \\ 
PG\,$1653{+}131$ & $25.4$ & $5.41$ & $-2.70$ & $ 8.3 \pm 1.0$ & {\citet{2013A&A...557A.122G}} \\ 
PG\,$0342{+}026$ & $26.0$ & $5.59$ & $-2.69$ & $6.2 \pm 1.0$ & {\citet{2013A&A...557A.122G}} \\ 
GD\,$108$ & $26.1$ & $5.58$ & $-3.46$ & $6.0 \pm 1.8$ & {\citet{2013A&A...557A.122G}} \\ 
Feige\,$65$ & $26.2$ & $5.31$ & $-2.75$ & $7.2 \pm 1.1$ & {\citet{2013A&A...557A.122G}} \\ 
PHL\,$44$ & $26.6$ & $5.41$ & $-2.97$ & $8.4 \pm 1.0$ & {\citet{2013A&A...557A.122G}} \\ 
SB\,$815$ & $27.0$ & $5.32$ & $-2.90$ & $ 7.3 \pm 1.0$ & {\citet{2013A&A...557A.122G}} \\ 
PG\,$2205{+}023$ & $27.1$ & $5.51$ & $<-4.00$ & $<10.0$ & {\citet{2013A&A...557A.122G}} \\ 
PG\,$2314{+}076$ & $27.2$ & $5.65$ & $<-4.00$ & $6.0 \pm 2.2$ & {\citet{2013A&A...557A.122G}} \\ 
EC\,$03591{-}3232$ & $28.0$ & $5.55$ & $-2.03$ & $4.8 \pm 1.0$ &  {\citet{2013A&A...557A.122G}} \\ 
EC\,$12234{-}2607$ & $28.0$ & $5.58$ & $-1.58$ & $ 6.8 \pm 1.4$ & {\citet{2013A&A...557A.122G}} \\ 
PG\,$2349{+}002$ & $28.0$ & $5.73$ & $-3.45$ & $ 5.7 \pm 1.5$ & {\citet{2013A&A...557A.122G}} \\ 
EC\,$01120{-}5259$ & $28.9$ & $5.41$ & $-2.54$ & $ 5.8 \pm 1.2$ & {\citet{2013A&A...557A.122G}} \\ 
EC\,$03263{-}6403$ & $29.3$ & $5.48$ & $-2.51$ & $<5.0$ & {\citet{2013A&A...557A.122G}} \\ 
PG\,$1303{+}097$ & $29.8$ & $5.83$ & $-2.17$ & $6.1 \pm 1.5$ & {\citet{2013A&A...557A.122G}} \\ 
EC\,$03470{-}5039$ & $30.5$ & $5.61$ & $<-4.00$ & $ 7.3 \pm 2.0$ & {\citet{2013A&A...557A.122G}} \\ 
PG\,$1710{+}490$ & $30.6$ & $5.66$ & $-2.43$ & $7.1 \pm 1.6$ & {\citet{2013A&A...557A.122G}} \\ 
Feige\,$38$ & $30.6$ & $5.83$ & $-2.37$ & $5.3 \pm 1.0$ & {\citet{2013A&A...557A.122G}} \\ 
HE\,$0447{-}3654$ & $30.7$ & $5.57$ & $<-3.00$ & $7.3 \pm 1.8$ & {\citet{2013A&A...557A.122G}} \\ 
EC\,$22081{-}1916$ & $31.1$ & $4.77$ & $-1.97$ & $163 \pm 3.0$ & {\citet{2013A&A...557A.122G}} \\ 
EC\,$14248{-}2647$ & $31.4$ & $5.56$ & $-1.64$ & $7.0 \pm 1.5$ & {\citet{2013A&A...557A.122G}} \\ 
EC\,$02542{-}3019$ & $31.9$ & $5.68$ & $-1.89$ & $ 7.3 \pm 1.5$ & {\citet{2013A&A...557A.122G}} \\ 
EC\,$21043{-}4017$ & $32.4$ & $5.63$ & $-1.58$ & $5.6 \pm 1.8$ & {\citet{2013A&A...557A.122G}} \\ 
EC\,$20229{-}3716$ & $32.5$ & $5.00$ & $-1.75$ & $4.5 \pm 1.0$ & {\citet{2013A&A...557A.122G}} \\ 
EC\,$05479{-}5818$ & $33.0$ & $5.93$ & $-1.66$ & $5.8 \pm 1.1$ & {\citet{2013A&A...557A.122G}} \\ 
EC\,$13047{-}3049$ & $34.7$ & $5.35$ & $-2.57$ & $ 6.8 \pm 3.6$ & {\citet{2013A&A...557A.122G}} \\ 
{[}CW83{]}\,$1758{+}36$ & $34.6$ & $5.79$ & $-1.51$ & $5.7 \pm 1.4$ &  {\citet{2013A&A...557A.122G}} \\ 
PHL\,$334$ & $34.8$ & $5.84$ & $-1.42$ & $<5.0$ & {\citet{2013A&A...557A.122G}} \\ 
PG\,$0909{+}164$ & $35.3$ & $5.33$ & $-2.76$ & $<10.0$ & {\citet{2013A&A...557A.122G}} \\ 
PG\,$0909{+}276$ & $35.5$ & $6.09$ & $-1.00$ & $ 9.3 \pm 1.4$ & {\citet{2013A&A...557A.122G}} \\ 
EC\,$03408{-}1315$ & $35.7$ & $5.85$ & $-1.61$ & $8.8 \pm 1.8$ & {\citet{2013A&A...557A.122G}} \\ 
PG\,$1505{+}074$ & $37.1$ & $5.39$ & $-2.69$ & $<5.0$ & {\citet{2013A&A...557A.122G}} \\ 
PG\,$1616{+}144$ & $37.3$ & $5.95$ & $-1.26$ & $<10.0$ & {\citet{2013A&A...557A.122G}} \\ 
PHL\,$1548$ & $37.4$ & $5.79$ & $-1.55$ & $ 9.1 \pm 1.6$ & {\citet{2013A&A...557A.122G}} \\ 
{[}CW83{]}\,$0512{-}08$ & $38.4$ & $5.77$ & $-0.73$ & $7.7 \pm 1.1$ &  {\citet{2013A&A...557A.122G}} \\ 
PB\,$5333$ & $40.6$ & $5.96$ & $-2.62$ & $<10.0$ & {\citet{2013A&A...557A.122G}} \\ 
SB\,$290$ & $26.3$ & $5.31$ & $-2.52$ & $58 \pm 1$ & {\citet{2013A&A...551L...4G}} \\ 
PG\,$1047{+}003$ & $33.1$ & $5.80$ &  &  & {\citet{2012A&A...539A..12F}} \\ 
PG\,$1219{+}534$ & $33.6$ & $5.81$ & $-1.49$ & $< 2.1$ & {\citet{2012A&A...539A..12F}} \\ 
EC\,$05217{-}3914$ & $32.0$ & $5.73$ &  &  & {\citet{2012A&A...539A..12F}} \\ 
PG\,$1325{+}101$ & $35.0$ & $5.81$ & $-1.70$ & $< 30.0$ & {\citet{2012A&A...539A..12F}} \\ 
PG\,$0911{+}456$ & $31.9$ & $5.78$ & $-2.55$ & & {\citet{2012A&A...539A..12F}} \\ 
BAL\,$090100001$ & $28.0$ & $5.38$ & $<-2.50$ & & {\citet{2012A&A...539A..12F}} \\ 
EC\,$09582{-}1137$ & $34.8$ & $5.79$ & $-1.68$ & & {\citet{2012A&A...539A..12F}} \\ 
KPD\,$1943{+}4058$ & $27.7$ & $5.52$ & $-2.71$ & & {\citet{2012A&A...539A..12F}} \\ 
KPD\,$0629{-}0016$ & $26.5$ & $5.45$ & $-2.79$ & & {\citet{2012A&A...539A..12F}} \\ 
KIC$02697388$ & $25.4$ & $5.49$ & $-2.67$ & & {\citet{2012A&A...539A..12F}} \\
\bottomrule
\end{tabular}
\end{minipage}
\end{table*}

\label{lastpage}
\end{document}